\documentclass[conference]{IEEEtran}
\usepackage{graphicx}
\usepackage{amsmath}
\usepackage{lipsum}
\usepackage{amsthm}
\usepackage[strings]{underscore}
\usepackage{amsfonts}
\usepackage{caption}
\usepackage{siunitx}
\usepackage{amssymb}
\usepackage{subcaption}
\usepackage{placeins}
\usepackage{lettrine}
\usepackage{fancyhdr} 
\usepackage{multicol, blindtext}
\usepackage{bbm}
\usepackage{calrsfs}
\usepackage{cite}
\usepackage{color}
\usepackage{cuted}
\usepackage{mathtools}
\DeclarePairedDelimiter\ceil{\lceil}{\rceil}

\usepackage{multicol}
\usepackage{hyperref}
\usepackage{algorithm}
\usepackage{algpseudocode}
\usepackage{authblk}
\newtheorem{remark}{Remark}
\makeatletter

\newcommand{\Rmnum}[1]{\expandafter\@slowromancap\romannumeral #1@}

\newtheorem{theorem}{Theorem}

\newtheorem{proposition}{Proposition}
\newtheorem{definition}{Definition}
\newtheorem{lemma}{Lemma}
\makeatletter
\def\BState{\State\hskip-\ALG@thistlm}
\makeatother
\ifCLASSINFOpdf
\else
\fi

\begin{document}
\title{The Age of Incorrect Information: A New Performance Metric for Status Updates}
\author[*]{Ali Maatouk}
\author[*]{Saad Kriouile}
\author[*]{Mohamad Assaad}
\author[$\dagger$]{Anthony Ephremides}
\affil[*]{TCL Chair on 5G, Laboratoire des Signaux et Syst\`emes, CentraleSup\'elec, Gif-sur-Yvette, France }
\affil[$\dagger$]{ECE Dept., University of Maryland, College Park, MD 20742}
\maketitle
\thispagestyle{fancy}
\pagestyle{fancy}
\fancyhf{}
\fancyheadoffset{0cm}
\renewcommand{\headrulewidth}{0pt} 
\renewcommand{\footrulewidth}{0pt}
\renewcommand*{\thepage}{\scriptsize{\arabic{page}}}
\fancyhead[R]{\thepage}
\fancypagestyle{plain}{%
   \fancyhf{}%
   \fancyhead[R]{\thepage}%
}
 
\begin{abstract}
In this paper, we introduce a new performance
metric in the framework of status updates that we will refer
to as the Age of Incorrect Information (AoII). This new metric
deals with the shortcomings of both the Age of Information
(AoI) and the conventional error penalty functions as it neatly
extends the notion of fresh updates to that of fresh ``informative"
updates. The word informative in this context refers to updates
that bring new and correct information to the monitor side.
After properly motivating the new metric, and with the aim of
minimizing its average, we formulate a Markov Decision Process
(MDP) in a transmitter-receiver pair scenario where packets are
sent over an unreliable channel. We show that a simple ``always
update" policy minimizes the aforementioned average penalty
along with the average age and prediction error. We then tackle
the general, and more realistic case, where the transmitter cannot
surpass a specific power budget. The problem is formulated as
a Constrained Markov Decision Process (CMDP) for which we
provide a Lagrangian approach to solve. After characterizing the optimal transmission policy of the Lagrangian problem, we
provide a rigorous mathematical proof to showcase that a mixture
of two Lagrange policies is optimal for the CMDP in question.
Equipped with this, we provide a low complexity algorithm that finds the AoII-optimal
operating point of the system in the constrained scenario. Lastly,
simulation results are laid out to showcase the performance of the proposed policy and highlight the differences with the AoI
framework.
\let\thefootnote\relax\footnotetext{This work has been supported by the TCL chair on 5G, ONR N000141812046, NSF CCF1813078, NSF CNS1551040, and NSF CCF1420651.}
\end{abstract}

%Next, we show that, although optimized, the standard CSMA scheme still lacks behind other distributed schemes in terms of average age in some special cases. These results motivated us to propose a novel distributed scheme belonging to the CSMA family in the aim of improving the performance the optimized standard CSMA. To circumvent the intractability of finding a closed form of the total average age in our generalized proposed scheme, a Sequential Convex Approximation (\textbf{SCA}) approach is presented to optimize the age performance in the network. Simulations results are then laid out to highlight the performance gain offered by our SCA approach in comparison to the optimized standard CSMA.

%\begin{IEEEkeywords} 5G, Massive MIMO, Clustering, Two-Stage Beamforming, Scheduling, Low energy consumption, Wireless MAC. \end{IEEEkeywords}


% \fi
\IEEEpeerreviewmaketitle

\section{Introduction}
With the proliferation of cheap sensors and devices, monitoring has become the new standard of technology applications. In these applications, a monitor is interested in having accurate information about a remote process (e.g., a car's position and velocity \cite{5307471}, the humidity of a room\cite{5597912}, etc.). To achieve this goal, the transmitter side of the link sends time-stamped status updates over the network to maximize/minimize a specific performance metric. To address the shortcomings of the throughput and delay metrics in these types of scenarios, the Age of Information (\textbf{AoI}) has been introduced to capture the notion of information freshness. To that extent, the AoI quantifies the information time lag at the monitor. The motivation for such
a framework is that having fresh knowledge about the process
of interest should result in a better real-time estimation
of the process. This prompted a surge of papers on the subject to explore the potentials of the metric mentioned above. Consequently, the AoI is now widely regarded as a fundamental performance measure in communication systems.

Since its establishment in \cite{6195689}, the efforts of researchers in the AoI area were divided on a wide variety of real-life scenarios that arise in communication networks. For example, the AoI was heavily studied in the framework of energy harvesting sources in \cite{2018arXiv180202129A}. Optimizing the average age in the case where the transmitter generates packets at will was considered in \cite{8000687} where, interestingly, it was shown that a zero-wait policy is far from being optimal. The AoI metric has also been recently used as a performance metric in content caching \cite{8006506}. Centralized scheduling with the goal of minimizing the average age has also captured a lot of research attention (e.g., \cite{8006590,2018arXiv180704356Z,8845254,2020arXiv200103096M}). For instance, the optimization of the average AoI with hybrid ARQ under a resource constraint was studied in \cite{8377368}. \color{black}In another line of work, and as the AoI is of wide interest in sensors applications where devices are autonomous, distributed scheduling schemes were proposed in \cite{2018arXiv180306469T,2018arXiv180103975J,2019arXiv190100481M,9007478}. For example, age-optimal back-off timers in CSMA environments were found in \cite{9007478}. Since streams normally have different priority assignments, researchers have lately focused on studying the AoI in multi-class scenarios \cite{2018arXiv180805738Z,2018arXiv180104067N,8613408,8849695,2020arXiv200201916M}.

As seen above, most of the research in the AoI area has been heavily focused on calculating and optimizing the average AoI. However, as previously stated, the ultimate goal in the communication system in question is to have the best real-time remote estimation of the process of interest at the monitor side. This leads to the following important question: is the AoI really the perfect metric to be used to estimate in real-time a process remotely? There have been some recent efforts to try and answer this question. For example, it was shown in \cite{8006542} that the optimal minimum Mean Squared Error (\textbf{MSE}) policy of a Wiener process over a channel with random delay is far from being age-optimal. This stems from the fact that
the AoI, by definition, does not capture well the information
content of the transmitted packets nor the current
knowledge at the monitor. In fact, even when the monitor has
perfect knowledge of the process in question, the AoI always
increases with time and, therefore, an unnecessary penalty is paid. This basic observation showcases why the AoI may come short in this type of application. Similarly, the AoI was shown in \cite{2018arXiv181205215J} to be sub-optimal in minimizing the status error in remotely estimating  Markovian sources. These observations prompted efforts to propose new performance metrics that deal with the shortcomings of the AoI. Among these efforts, a time-based metric dubbed as the Age of Synchronization (\textbf{AoS}) was introduced in the framework of content caching \cite{8437927}. Specifically, the AoS is zero when the transmitter has no packets to send and it grows linearly with time when the transmitter side generates a new packet. Although the AoS includes the packets generation as a factor, it does not take into account the information structure of the source and the current estimate at the receiver, which limits its usage in remote estimation applications. In another work \cite{8406891}, the authors proposed different \emph{effective} age metrics for which a lower effective age should undoubtedly lead to a lower prediction
error. For example, the notion of \emph{Sampling Age} was introduced and was defined as the age relative to an ideal sampling pattern $g(t)$ that minimizes the error. However, finding the optimal pattern $g(t)$ was deemed to be far from being trivial. \color{black} As seen from the above efforts, the ultimate goal has been to propose new metrics or sampling/scheduling policies that minimize either the prediction error or the MSE. This raises a question of paramount importance: should the minimization of prediction error or mean squared error always be regarded as the definitive goal of the remote estimation scenario? To argue that this should not always be the case, we shed light on one of the shortcomings of these conventional error measures. The primary issue with these error functions is that they do not increasingly penalize the monitor for wrongfully estimating the process of interest. In other words, the same penalty is paid for being in an erroneous state no matter how long the monitor has been in it. To that extent, a monitor wrongfully thinking that a machine is at a normal temperature suffers from the same penalty no matter how long the machine has been overheating for. This clearly suggests that a more general framework should be introduced to deal with the shortcomings of these error measures.

\color{black}
In our paper, we pave the way for such a framework by introducing a new performance metric that deals with the above mentioned shortcomings of both the AoI and the error functions. To that end, we summarize in the following the key contributions of this paper:
\begin{itemize}
\item We first go into more depth on highlighting the shortcomings of the AoI and the error performance metrics in the case of remote process estimation. Aiming to deal with these shortcomings, we propose a new performance measure, which we will call the Age of Incorrect Information (\textbf{AoII}), that neatly extends the notion of fresh updates to that of fresh ``informative" updates. The word informative refers to updates that bring new and correct information to the monitor side. This new measure also captures the deteriorating effect the wrong information can have with time on the system.\color{black}
\item Afterward, we focus on the case where a transmitter-receiver pair communicates over an unreliable channel. The transmitter sends status updates about an $N$ states Markovian information source with the goal of the receiver being to estimate it accurately. In this scenario, we aim to find the optimal transmission policy that minimizes the average proposed metric. By casting this problem into a Markov Decision Process (\textbf{MDP}), we show that in the case where no constraints on the power are imposed, an ``always update" policy is able to minimize the average age, the prediction error, and the average AoII.
\item Following that, we tackle the more realistic case where each transmission incurs a cost, and the transmitter has a power budget that cannot be surpassed. We cast our problem in this case into a Constrained Markov Decision Process (\textbf{CMDP}) that is known to be challenging to solve. To circumvent this difficulty, we provide a Lagrange approach that transforms the CMDP to an unconstrained MDP. The Lagrangian optimization problem is then thoroughly studied and structural results on its optimal policy are provided. 
\item Subsequently, we provide a rigorous mathematical proof to show that the optimal operating point of the CMDP is achieved by a mixture of two deterministic Lagrange policies. Similar results were established in the literature for the AoI optimization framework \cite{8377368}. However, due to the inherent properties of the proposed AoII metric, the standard approach adopted in \cite{8377368} cannot be followed. Specifically, the mathematical expressions involved in our case are not necessarily convex, which limits the applicability of the approach in \cite{8377368}. Accordingly, we proceed in a different direction to establish the required results as will be seen in later sections of the paper. Armed with these results, we provide an algorithm that finds the AoII-optimal policy under the power constraint in logarithmic complexity. 
\color{black}
\item Lastly, we provide numerical implementations of our transmission policy that highlight its performance and showcase interesting insights on the differences between the AoI and the AoII frameworks.
\end{itemize}
The rest of the paper is organized as follows: Section II
is dedicated to the motivation of the newly proposed framework. The system model, along with the dynamics of the proposed metric are presented in Section III. Section IV provides the MDP description of the problem along with its analysis in the unconstrained power scenario. In Section V, we thoroughly analyze the constrained scenario and propose an optimal approach to solve it. Numerical results that corroborate the theoretical findings are laid out in Section VI, while the paper is concluded in Section VII.
\begin{figure*}[ht]
\centering
\begin{subfigure}{0.33\textwidth}
  \centering
  \includegraphics[width=.99\linewidth]{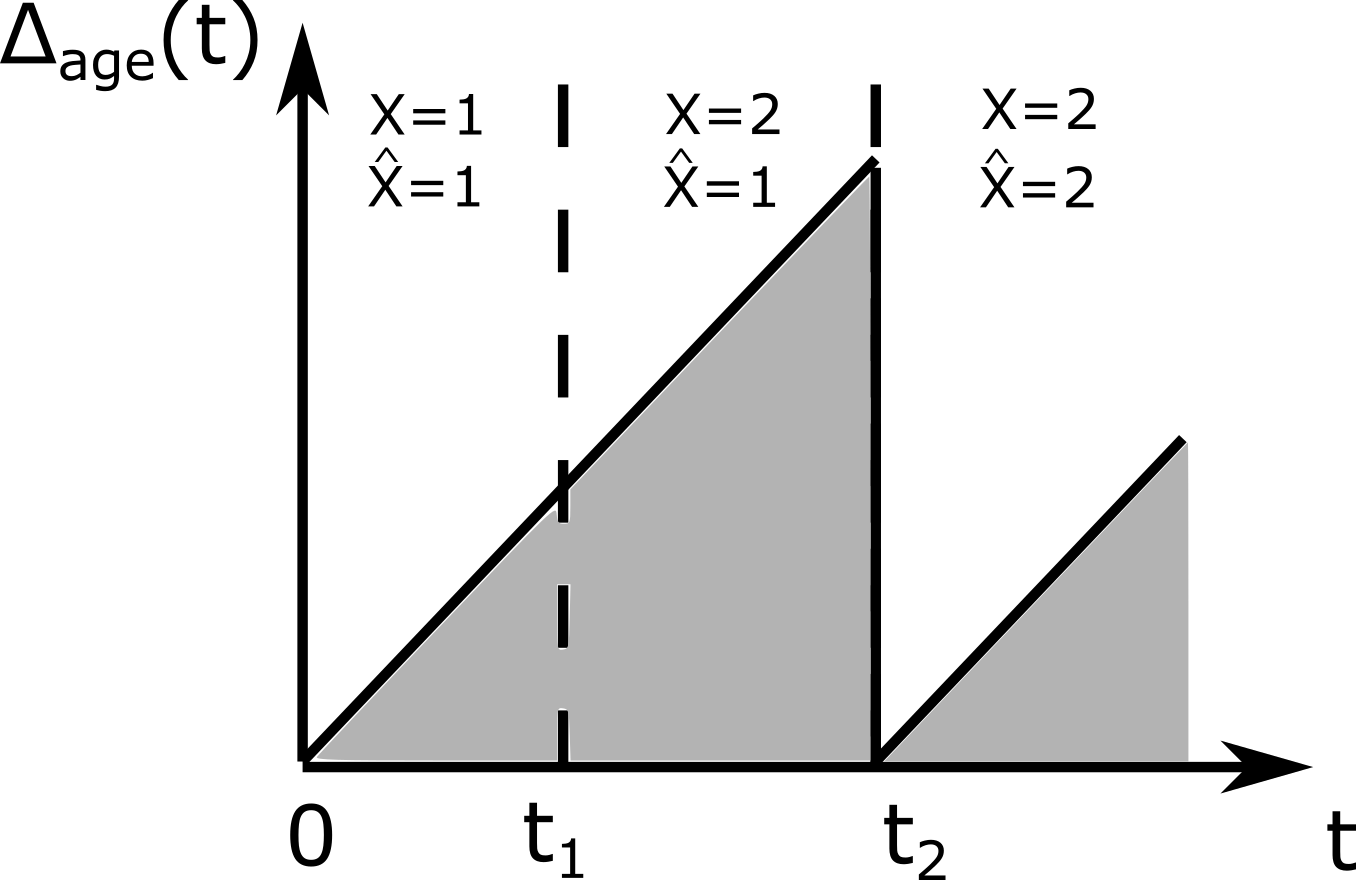}
  \caption{Age penalty function.}
    \label{agemetric}
\end{subfigure}%    
\begin{subfigure}{0.33\textwidth}
\centering
  \includegraphics[width=.99\linewidth]{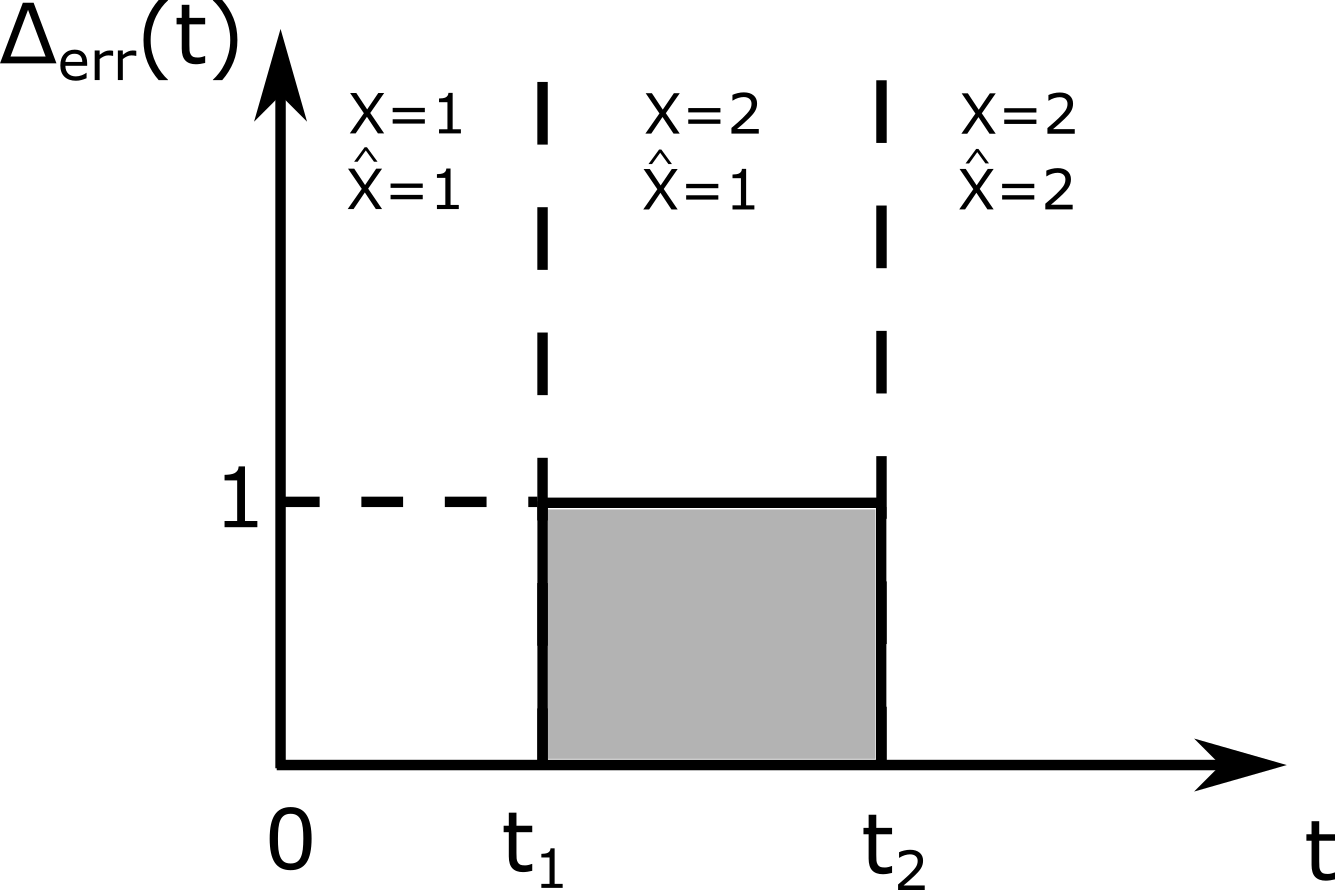}
  \caption{Error penalty function.}
\label{errmetric}
\end{subfigure}%
\begin{subfigure}{0.33\textwidth}
\centering
  \includegraphics[width=.99\linewidth]{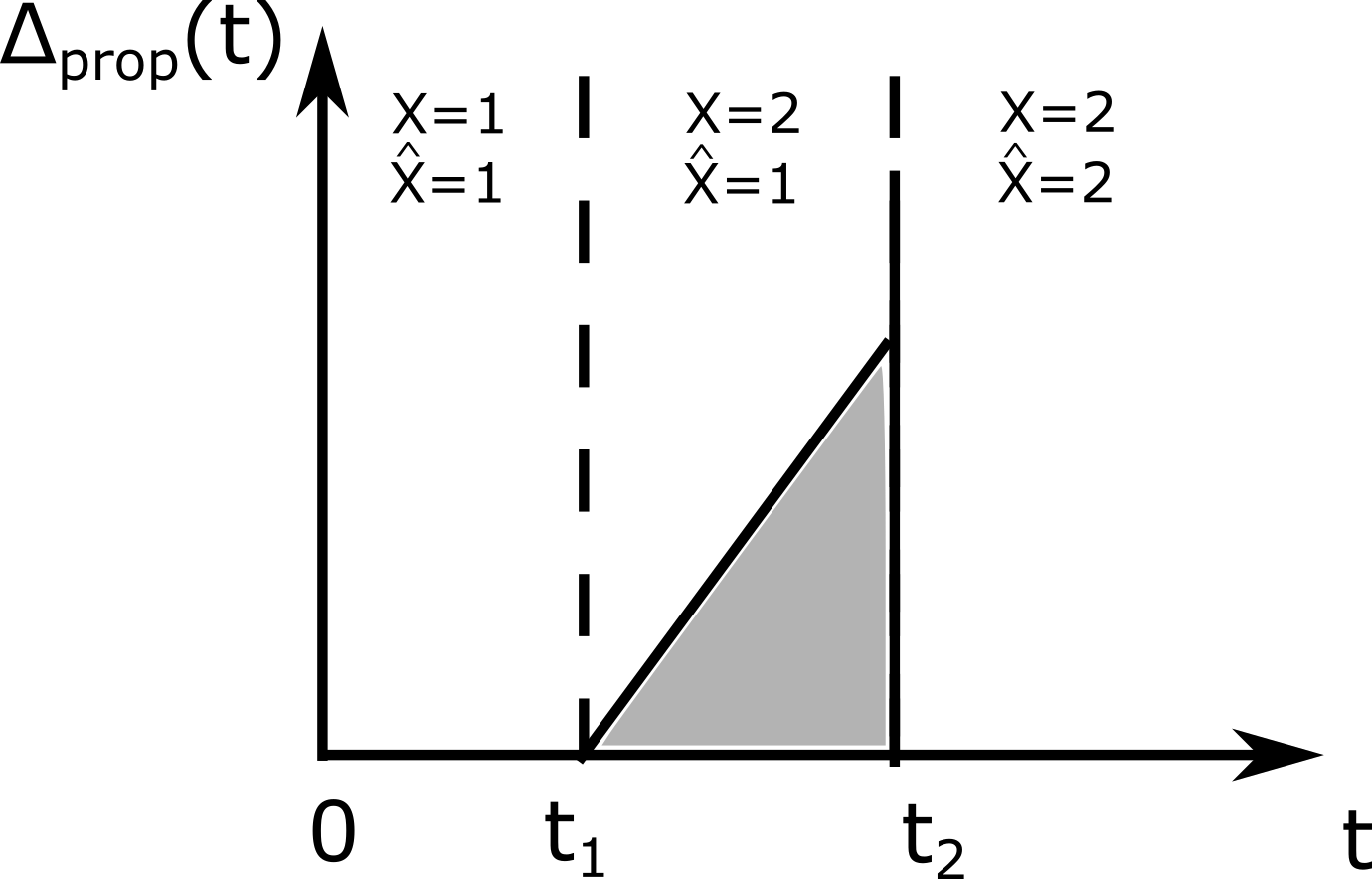}
  \caption{Proposed penalty function.}
\label{effmetric}
\end{subfigure}%
\caption{Illustrations of different penalty functions.}
\vspace{-22pt}
\label{metrics}
\end{figure*}
\section{Proposed metric}
\begin{figure}[!ht]
\centering
\includegraphics[width=.8\linewidth]{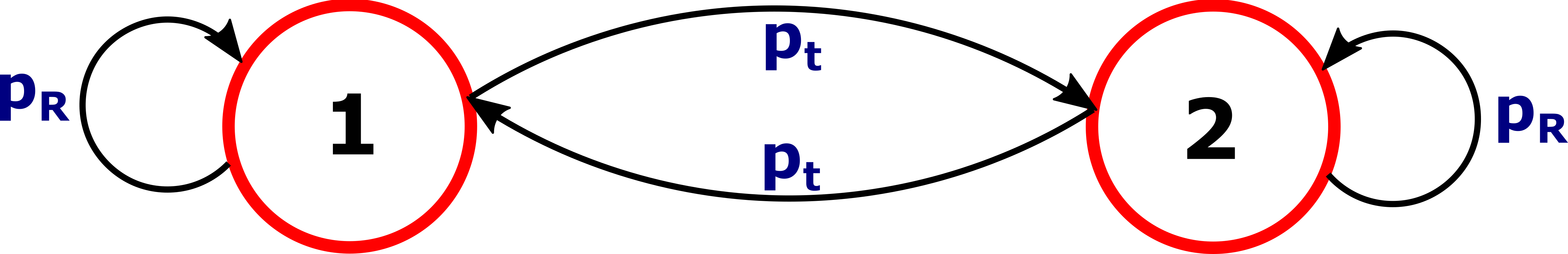}
\setlength{\belowcaptionskip}{-5pt}
\caption{Example of a two states information source.}
\label{sourceexample}
\end{figure}

To put into perspective our line of work, we focus in this section on a particular scenario where a transmitter-receiver pair communicates. More specifically, the transmitter observes a process $X(t)$ and informs the receiver (monitor) about it by sending status updates over the network. Based on the last received update, the monitor constructs an estimate of the process, denoted by $\hat{X}(t)$. Time is considered to be discrete and normalized to the time slot duration. For simplicity, we suppose in this section that the process in question can only have two values $\{1,2\}$, as depicted in Fig \ref{sourceexample}. At each time slot, the probability of remaining in the same state is $p_R$ while the probability of transitioning to another state is $p_t$. The transmitter decides when to 
inform the monitor about the process $X(t)$ by adopting a transmission policy that aims to minimize the average of a particular penalty function. 

First, let us consider the age penalty function to more closely examine its shortcomings. To that extent, we define the age as follows:
\begin{equation}
\Delta_{\text{age}}(t)=t-U(t),
\label{agedef}
\end{equation}
where $U(t)$ is the time-stamp of the last successfully received packet by the monitor. Based on this definition, we can observe that the age captures the information \emph{time lag} at the monitor, in an attempt to achieve timely updates. As seen in (\ref{agedef}), the age always increases as time progresses regardless of the current information at the monitor, which makes it fall short in numerous applications. To see this, let us observe the trend of $\Delta_{\text{age}}(t)$ in the time interval $[0,t_1]$ of Fig. \ref{agemetric}. In this interval, the monitor has perfect knowledge of the process of interest $X(t)$ and, therefore, any new update received in this interval will not change the information currently available at the monitor. Regardless of that, we can clearly see that the age penalty keeps growing with time, i.e., a penalty is being paid for not being updated on the information process although the monitor currently has perfect knowledge of the process in question. This above observation clearly put into perspective the shortcoming of the age penalty function and let us emphasize on the fact that any relevant metric for the remote estimation of a process has to capture more meaningfully its
information content and the current knowledge at the receiver. 

Another widely used penalty function is the error penalty:
\begin{equation}
\Delta_{\text{err}}(t)=\mathbbm{1}{\{\hat{X}(t)\neq X(t)\}},
\label{errorfunc}
\end{equation}
where $\mathbbm{1}$ is the indicator function. In fact, minimizing the average of the function in (\ref{errorfunc}), is equivalent to the minimization of the prediction error $\Pr(\hat{X}(t)\neq X(t))$. The key shortcoming of this error penalty function is its failure to capture the following phenomena that arises in numerous applications: staying in an erroneous state should have an increasing penalty effect. In fact, the function in (\ref{errorfunc}) treats all instances of error equally, no matter how long the time elapsed since their start is. In other words, the penalty of being in an erroneous state after $1$ time slot, or $100$ time slots is the same value of $1$. Because of this observation, we can see that the long-time average error penalty due to a burst error is the same as the one resulting from several isolated errors of the same duration. However, this is not always the case, and there exists a vast amount of applications where the penalty grows the longer the monitor has incorrect information. \color{black} For example, let us suppose that $X(t)=1$ refers to the case where a machine is at a normal temperature at time $t$ while $X(t)=2$ is the case where the machine is overheating. This information has to be transferred to a monitor that can, consequently, react to the state of the machine. By considering the time interval $[t_1,t_2]$ of Fig. \ref{errmetric}, we can see that no matter how long the duration of the interval $\Delta t=t_2-t_1$, the same penalty $\Delta_{err}(t)=1$ is kept. However, as it is well-known, the repercussions of keeping a machine overheated become more severe as time goes on. Therefore, this should be reflected in the adopted penalty function and should be considered as one of its key design features. It is worth mentioning that the list of such real life applications, where the level of dissatisfaction grows as time progresses, is vast. We report a few examples in the following:
\begin{itemize}
\item A real-time video stream in which packets are sent through a channel, and where losses can occur due to, for example, an inaccurate channel estimate. Similarly to the previous case, the adoption of the AoI as a performance metric will fall short since a penalty is constantly paid even if the current channel estimate is accurate. On another note, if any standard error penalty function is adopted, the effect of burst errors on the performance is not captured. However, it is well-known that in this application, a burst packet losses lead to more distortion of the video when compared to an equal number of isolated losses. 
\item An actuator that can tolerate inaccurate actions for a brief amount of time. However, when these actions are done for a long duration, substantial performance penalties are to be paid.
\item The relay of fire outbreaks in environmental monitoring applications where any relay failure cause more severe repercussions the longer it lasts.
\end{itemize}
\color{black}
Motivated by all this, we aim to propose in our paper a new metric that elegantly combines the following two characteristics of the age and the error penalty functions:
\begin{enumerate}
\item The proposed metric captures the information content of the updates and the current knowledge of the monitor as done by the error penalty function in (\ref{errorfunc}).
\item The proposed metric captures the increasing dissatisfaction with time that is offered by the age penalty.
\end{enumerate}
Based on this, the general metric that we are about to introduce can be thought to capture the notion of \textbf{\emph{fresh informative}} updates. The word informative in this context refers to updates that bring \textbf{\emph{new}} information to the monitor side. In other words, when the monitor already has perfect knowledge about the process in question, we should not pay any penalty. However, as the state of the process change and the monitor becomes in an erroneous state, an update from the transmitter becomes informative. Because we need this update to arrive as \textbf{\emph{fresh}} as possible, we let the penalty grows with time as long as we are in an erroneous state. To that extent, our proposed metric, which we will call the Age of Incorrect Information (\textbf{AoII}), can be written as follows:
\begin{equation}
\Delta_{\text{AoII}}(t)=f(t)\times g(X(t),\hat{X}(t)),
\label{propfunc}
\end{equation}
where $f(t)$ is an increasing time penalty function, paid for being unaware of the correct status of the process for a certain amount of time. On the other hand, $g(X(t),\hat{X}(t))$ is an information penalty function that reflects the difference between the current estimate at the monitor and the actual state of the process. There exists a wide variety of choices for $f$ and $g$ that we can pick from. We list below some of these examples, starting with $g$ and following it by $f$.
\begin{itemize}
\item The indicator error function:
\begin{equation}
g_{\text{ind}}(X(t),\hat{X}(t))=\mathbbm{1}_{\{X_t\neq\hat{X}_t\}}.
\label{functional7}
\end{equation}
This information penalty function can be adopted when any mismatch between $X(t)$ and $\hat{X}(t)$ penalizes the system in the same fashion. 
\item The squared error function:
\begin{equation}
g_{\text{sq}}(X(t),\hat{X}(t))=(X(t)-\hat{X}(t))^2.
\label{functional8}
\end{equation}
Unlike $g_{\text{ind}}(X(t),\hat{X}(t))$, this information penalty function penalizes more the system the larger the difference between $X(t)$ and $\hat{X}(t)$ is.
\item The threshold error function:
\begin{equation}
g_{\text{threshold}}(X(t),\hat{X}(t))=\mathbbm{1}_{|X(t)-\hat{X}(t)|\geq c},
\label{functional9}
\end{equation}
where $c>0$ is a predefined threshold. This information penalty function can be used when the system can tolerate small mismatches between $X(t)$ and $\hat{X}(t)$. However, when the mismatch between the two is high, a penalty is paid. 
\end{itemize}
Next, we provide examples of the time-dissatisfaction function $f$. To do so, we first define $V(t)$ as the last time instant where $g(X(t),\hat{X}(t))$ was equal to $0$. In other words, $V(t)$ is the last time instant where the monitor had zero information penalty, i.e., when the monitor had accurate information about the source. By leveraging this notion, we present the following examples of $f$.
\begin{itemize}
\item The linear time-dissatisfaction function:
\begin{equation}
f_{\text{linear}}(t)=t-V(t).
\label{functional1}
\end{equation}
\item The exponential time-dissatisfaction function:
\begin{equation}
f_{\text{exponential}}(t)=\exp(a(t-V(t))),
\label{functional2}
\end{equation}
where $a>0$ is a positive constant. This time-dissatisfaction function can be used when the system is extremely vulnerable to wrong information and the need for fresh correct information grows quickly with time. 
\item The time-threshold dissatisfaction function:
\begin{equation}
f_{\text{threshold}}(t)=\mathbbm{1}_{\{t-V(t)\geq d\}},
\label{functional3}
\end{equation}
where $\mathbbm{1}_{\{.\}}$ is the indicator function, and $d>0$ is a fixed time threshold that should not be violated. This time-dissatisfaction function can be adopted when the system's performance starts deteriorating due to wrong information beyond a certain time duration $d$.
\end{itemize}
\color{black}
For simplicity, we focus in the sequel on the case where  $f(t)=f_{\text{linear}}$ and $g(X(t),\hat{X}(t))=g_{\text{ind}}$. Specifically, we have:
\begin{equation}
\Delta_{prop}(t)=f(t)\times g(X(t),\hat{X}(t))=(t-V(t))\mathbbm{1}{\{\hat{X}(t)\neq X(t)\}}.
\label{expressionob}
\end{equation}
A sketch of this function is given in Fig. \ref{effmetric} where we can see how the penalty increases as time progresses in the interval $[t_1,t_2]$ to reflect the increasing dissatisfaction of being in an erroneous state. This metric will be the basis of our analysis in the upcoming sections, where we aim to minimize its average in a general scenario of interest. With that in mind, we stress the fact that our proposed metric is far more general and is not limited to this choice of $f$ and $g$.
\color{black}
\section{System Overview}
\subsection{System Model}
\begin{figure}[!ht]
\centering
\includegraphics[width=.65\linewidth]{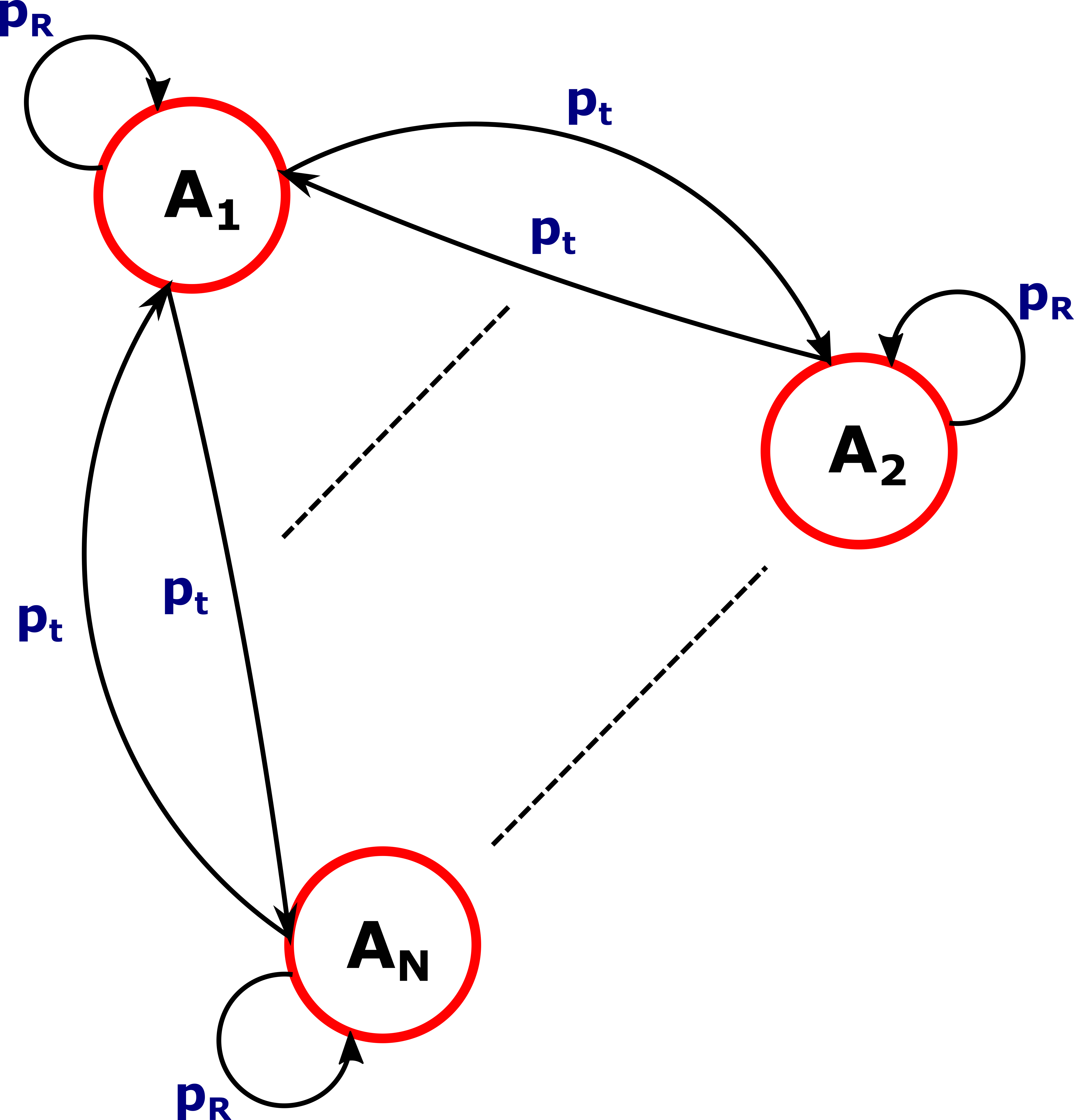}
\setlength{\belowcaptionskip}{-12pt}
\caption{Illustration of the process of interest.}
\label{sourcechain}
\end{figure}

We consider in our paper a transmitter-receiver pair where the transmitter sends status updates about the process of interest to the receiver side over an unreliable channel. Time is considered to be slotted and normalized to the slot duration (i.e., the slot duration is taken as $1$). The information process of interest is an $N$ states discrete Markov chain $\big(X(t)\big)_{t\in\mathbb{N}}$ depicted in Fig. \ref{sourcechain}. To that extent, we define the probability of remaining at the same state in the next time echelon as $\Pr(X(t+1)=X(t))=p_R$. Similarly, the probability of transitioning to another state is defined as $\Pr(X(t+1)\neq X(t))=p_t$. Since the process in question can have one of $N$ different possible values, the following always holds:
\begin{equation}
p_R+(N-1)p_t=1.
\end{equation}
\vspace{-20pt}

As for the unreliable channel model, we suppose that the channel realizations are independent and identically distributed (i.i.d.) over the time slots and follow a Bernoulli distribution. More precisely, the channel realization $h(t)$ is equal to $1$ if the packet is successfully decoded by the receiver side and is $0$ otherwise. To that extent, we define the success probability as $\Pr(h(t)=1)=p_s$ and the failure probability as $\Pr(h(t)=0)=p_f=1-p_s$. We consider that when a packet is delivered to the receiver, the receiver sends an Acknowledgement (\textbf{ACK}) packet back to the transmitter. In the case of a failure of transmission, a negative-acknowledgment (\textbf{NACK}) is sent by the receiver. We suppose that the ACK/NACK packets are instantaneously delivered to the transmitter \cite{8000687,2018arXiv180101803K}. This assumption is widely used in the literature since the ACK/NACK packets are small and, accordingly, their transmission times can be considered to be negligible. Using these ACK/NACK packets, the transmitter can have perfect knowledge of the information source estimate at the receiver at any time slot $t$. \color{black}

The next aspect of our model that we tackle is the nature of packets in the system. To that extent, we consider that the transmitter can generate information updates any time at its own will. More specifically, when the transmitter decides to send an update at time $t$, it samples the process $X(t)$ and proceeds to the transmission stage. If the packet is not successfully delivered to the receiver, and if the transmitter desires a transmission retrial at time $t+1$, a \emph{new} status update is generated by sampling $X(t+1)$ and the transmission stage begins again.

Lastly, and as previously explained in the preceding section, the transmitter's ultimate objective is to adopt a transmission policy that minimizes the time average of a particular penalty function. In the sequel, we adopt the newly proposed metric reported in (\ref{expressionob}). To fully characterize it, we provide details on its dynamics in the next subsection.
\subsection{Penalty Function Dynamics}
Let $S(t)$ be the penalty of the system mentioned above at time instant $t$. More specifically:
\begin{equation}
S(t)=(t-V(t))\mathbbm{1}{\{\hat{X}(t)\neq X(t)\}},
\label{expressionofs}
\end{equation}
where $V(t)$ is the last time instant where the monitor was in a correct state. In the sequel, we provide details concerning the dynamics of $S(t)$ in the aim of characterizing the values of $S(t+1)$. To do so, we first define $\psi(t)$ as the decision at time $t$ of the transmitter to either transmit (value $1$) or remain idle (value $0$). We distinguish in the following between two cases: $S(t)=0$ and $S(t)\neq0$.
\subsubsection{$S(t)=0$} In this case, the monitor has perfect knowledge of the process of interest at time $t$. If the transmitter decides not to send a status update, then $S(t+1)$ will be equal to $0$ if the process does not change value. This happens with a probability $p_R$. In the same fashion, $S(t+1)$ will be equal to $1$ if the process changes value, which happens with a probability $1-p_R=(N-1)p_t$. Let us now consider the case where the transmitter decides to send a status update at time $t$. Regardless of the channel realization, no new information will be conveyed to the monitor as $\hat{X}(t+1)$ will have the same value of $\hat{X}(t)$. Consequently, the previous analysis still holds for this case and $S(t+1)$ will be equal to $0$ if the process does not change value and $1$ otherwise. We summarize what was stated in the following:
\begin{itemize}
\item $\Pr\big(S(t+1)=0|S(t)=0,\psi(t)=0\big)=\Pr\big(S(t+1)=0|S(t)=0,\psi(t)=1\big)=p_R$
\item $\Pr\big(S(t+1)=1|S(t)=0,\psi(t)=0\big)=\Pr\big(S(t+1)=1|S(t)=0,\psi(t)=1\big)=1-p_R=(N-1)p_t$
\end{itemize} 
\subsubsection{$S(t)\neq0$} In this case, the monitor does not have correct knowledge of the process of interest (i.e., $\hat{X}(t)\neq X(t)$ ). If the transmitter decides to remain idle, then $S(t+1)$ will be equal to $0$ if and only if the information process changes to the value that the monitor has from its last received update. More specifically, this is when $X(t+1)=X(U(t))$ with $U(t)$ being the time-stamp of the last successfully received packet by the monitor. This event occurs with a probability $p_t$. On the other hand, if the process keeps its same value, or transition to one of the remaining $N-2$ states, the penalty will grow by a step, i.e., $S(t+1)=S(t)+1$. Now, let us consider the case where the transmitter decides to send a packet. To that extent, we consider two cases:
\begin{itemize}
\item $h(t)=0$: In this case, the transmitted packet is not successfully decoded by the receiver. Therefore, no new knowledge is given to the monitor, i.e., $\hat{X}(t+1)=\hat{X}(t)$. To that extent, conditioned on $h(t)=0$, we can assert that $S(t+1)$ becomes zero if and only
if the information process changes to the value that the monitor has from its last received update. As previously mentioned, this event occurs with a probability $p_t$. On the other hand, $S(t+1)$ will be equal $S(t)+1$ if the process keeps its same value or change to one of the other $N-2$ states, which happens with a probability $p_R+(N-2)p_t$.
\item $h(t)=1$: In this case, the transmitted packet is successfully decoded by the receiver. Therefore, the estimate at the monitor $\hat{X}(t+1)$ is nothing but $X(t)$. To that extent, $S(t+1)$ will be equal to zero if the information process did not change during the transmission slot. This event happens with a probability $p_R$. On the other hand, if the process has changed during transmission to any of the remaining $N-1$ states, $S(t+1)$ will increase by $1$.
\end{itemize}
By taking into account the independence between the information process transitions and the channel realizations, we can summarize the transitions probabilities of $S(t)$ in the following:
\begin{itemize}
\item $\Pr\big(S(t+1)=0|S(t)\neq0,\psi(t)=0\big)=p_t$
\item $\Pr\big(S(t+1)=S(t)+1|S(t)\neq0,\psi(t)=0\big)=p_R+(N-2)p_t$
\item $\Pr\big(S(t+1)=0|S(t)\neq0,\psi(t)=1\big)=p_Rp_s+p_fp_t$
\item $\Pr\big(S(t+1)=S(t)+1|S(t)\neq0,\psi(t)=1\big)=p_Rp_f+(N-2)p_t+p_sp_t$
\end{itemize}
\section{Unconstrained Scenario}
\subsection{Problem Formulation}
The objective of this paper is to find a transmission policy that minimizes the total average AoII of the network. A transmission policy $\phi$ is defined as a sequence of actions $\phi=(\psi^{\phi}(0),\psi^{\phi}(1),\ldots)$ where $\psi^{\phi}(t)=1$ if a transmission is initiated at time $t$. By letting $\Phi$ denote the set of all possible causal scheduling policies, our problem can be formulated as follows:
\begin{equation}
\setlength{\belowdisplayskip}{0pt} \setlength{\belowdisplayshortskip}{0pt}
\setlength{\abovedisplayskip}{0pt} \setlength{\abovedisplayshortskip}{0pt} 
\begin{aligned}
& \underset{\phi\in \Phi}{\text{minimize}}
& & \lim_{T\to+\infty} \text{sup}\:\frac{1}{T}\mathbb{E}^{\phi}\Big(\sum_{t=0}^{T-1}S^{\phi}(t)|S(0)\Big).
\end{aligned}
\label{originalobjective}
\end{equation}
\subsection{MDP Characterization}
Based on our model's assumptions and the dynamics previously detailed in Section III-B, our problem in (\ref{originalobjective}) can be cast into an infinite horizon average cost Markov decision process that is defined as follows:
\begin{itemize}
\item \textbf{States}: The state of the MDP at time $t$ is nothing but the penalty function $S(t)$. This penalty can have any value in $\mathbb{N}$. Therefore, the considered state space is countable and infinite.
\item \textbf{Actions}: The action at time $t$, denoted by $\psi(t)$, indicates if a transmission is attempted (value $1$) or the transmitter remains idle (value $0$).
\item \textbf{Transitions probabilities}: The transitions probabilities between the different states have been previously detailed in Section III-B.
\item \textbf{Cost}: We let the instantaneous cost of the MDP, $C(S(t),\psi(t))$, to be simply the penalty function $S(t)$.
\end{itemize}
Finding the optimal solution of an infinite horizon average cost MDP is recognized to be challenging due to the curse
of dimensionality. More precisely, it is well-known that the optimal policy $\phi^*$ of the problem mentioned above can be obtained by solving the following Bellman equation \cite{Bertsekas:2000:DPO:517430}:
\begin{equation}
\theta + V(S)=\min_{\psi\in\{0,1\}}\big\{S+\sum_{S'\in\mathbb{N} }\Pr(S\rightarrow S'|\psi)V(S')\big\} \quad \forall S\in\mathbb{N}
\label{bellman}
\end{equation}
where $\Pr(S\rightarrow S'|\psi)$ is the transition probability from state $S$ to $S'$ given the action $\psi$, $\theta$ is the optimal value of (\ref{originalobjective}) and $V(S)$ is the value function. Based on (\ref{bellman}), one can see that the optimal policy $\phi^*$ depends on $V(.)$, for which
there is no closed-form solution in general \cite{Bertsekas:2000:DPO:517430}. There exist various numerical algorithms in the literature that solve (\ref{bellman}), such as the value iteration and the policy iteration algorithms. However, they suffer from being computationally demanding. To circumvent this complexity, we study in the next section the structural properties of the optimal transmission policy. 
\subsection{Structural Results}
The first step in our structural analysis of the optimal policy consists of studying the particularity of the value function $V(.)$. To that extent, we provide the following lemma.
\begin{lemma}
The value function $V(S)$ is increasing in $S$.
\label{increasingvalue}
\end{lemma}
\begin{IEEEproof}
The proof can be found in Appendix \ref{prooflemma1} of the
supplementary material.
\end{IEEEproof}
The above lemma will be used in the following theorem to provide results on the optimal transmission policy.
\begin{theorem}
The optimal transmission policy $\phi^*$ of our problem in (\ref{originalobjective}) is:
\begin{itemize}
\item $p_t<p_R$: the transmitter should send updates at each time slot or when the receiver is in an erroneous state. In both cases, the optimal cost is:
\begin{equation}
\overline{C}_{AU}=(N-1)p_t\frac{\frac{1}{(1-a)^2}}{1+\frac{(N-1)p_t}{1-a}}.
\label{alwaysupdate}
\end{equation}
\item  $p_t\geq p_R$: it is optimal to never transmit any packet. In this case, the optimal cost is:
\begin{equation}
\overline{C}_{NU}=\frac{(N-1)p_t}{(1-b)^2+(1-b)(N-1)p_t}.
\label{noupdatescost}
\end{equation}
%\item if $p_t=p_R$, it does not matter. The cost is $\overline{C}_{AU}=\overline{C}_{NU}$
\end{itemize}
with $a,b$ being two constants that are equal to $p_Rp_f+(N-2)p_t+p_sp_t$ and $p_R+(N-2)p_t$ respectively.
\label{optimalpolicyunconstrained}
\end{theorem}
\begin{IEEEproof}
The proof can be found in Appendix \ref{prooftheorem1} of the
supplementary material.
%\subsubsection{$p_t=p_R$} In this scenario, the transmitter can either always transmit or never transmit and still achieve the same cost. Consequently, the optimal policy in this case is to simply choose any random action at each time slot $t$. 
\end{IEEEproof}
The intuition behind the above results is that when $p_t<p_R$, a transmitted packet has a high chance of becoming erroneous by the time it is delivered to the receiver. Accordingly, in this case, the information source changes so fast to the point that transmitting packets will harm the performance of the system. As this case is not of practical interest, we focus in the rest of the paper on the scenario where $p_t<p_R$. Consequently, we have that in the case where no constraints on the power are imposed, the optimal minimum cost is achieved either by sending updates at every time slot or when the receiver is in an erroneous state.  \color{black}
\begin{remark}
By adopting the same model as the one above, and by considering the AoI as the penalty function, we can verify by the same manners that the optimal transmission policy is to send updates at each time slot. As for the error penalty function, it can also be verified that sending updates at every time slot or when the receiver is in an erroneous case minimizes the prediction error. Consequently, and as the intuition suggests, an ``always update" policy minimizes all the above $3$ penalties in the unconstrained power case. However, as will be shown in the sequel, this does not hold in the case of power-constrained scenarios.
\end{remark}
%By using the results of the above Theorem, we can conclude that for any $n\in\mathbb{N}^{*}$, the optimal threshold is nothing but $n^*=1$. To fully characterize the general optimal policy for any $n\in\mathbb{N}$, we still have study the special case of always updating (i.e. $n=0$) in the following lemma. 
%\begin{lemma}
%The average cost incurred when the transmitter always sends updates is:
%\begin{equation}
%\overline{C}_{AU}=\overline{C}(1)=(N-1)p_t\frac{1+\frac{a(\frac{2-a}{1-a})}{1-a}}{1+(N-1)p_t+\frac{(N-1)p_ta}{1-a}}
%\label{costalwaysupdate}
%\end{equation}
%\end{lemma}
%\begin{IEEEproof}
%As it can be seen in the dynamics reported in Section III-B, the transitions from state $S=0$ do not depend on the action taken by the transmitter. Consequently, and by constructing the DTMC in the case where the transmitter always sends updates, it can be easily verified that the DTMC coincides with that of a threshold policy with $n=1$. Therefore, by replacing $n$ with $1$ in the average cost expression of (\ref{costgamma}), the proof can be concluded. 
%\end{IEEEproof}
%
%In this case, the following minimum cost is achieved:
%\begin{equation}
%\overline{C}_{\min}=(N-1)p_t\frac{1+\frac{a(\frac{2-a}{1-a})}{1-a}}{1+(N-1)p_t+\frac{(N-1)p_ta}{1-a}}
%\end{equation}
\section{Power constrained scenario}
\subsection{Problem Formulation}
In realistic scenarios, a transmitter cannot send status updates at each time slot. In fact, each attempted transmission incurs a power cost $\delta$, and the transmitter has an average power budget $\delta_{\text{budget}}$ that cannot be surpassed. Consequently, the transmitter has to choose wisely when to transmit an update to the monitor as the following constraint has to be satisfied by any chosen transmission policy $\phi$: 
\begin{equation}
\lim_{T\to+\infty} \text{sup}\:\frac{1}{T}\mathbb{E}^{\phi}\Big(\sum_{t=0}^{T-1}\delta\psi^{\phi}(t)\Big)\leq \delta_{\text{budget}},
\end{equation}
where the transmission policy $\phi$ is defined as a sequence of actions $\phi=(\psi^{\phi}(0),\psi^{\phi}(1),\ldots)$ such that $\psi^{\phi}(t)=1$ if a transmission is initiated at time $t$. Since $\lim_{T\to+\infty} \text{sup}\:\frac{1}{T}\mathbb{E}^{\phi}\Big(\sum_{t=0}^{T-1}\psi^{\phi}(t)\Big)\leq1$, we define $\alpha=\frac{\delta_{\text{budget}}}{\delta}$ and we suppose that $\alpha\leq1$ as the constraint becomes redundant otherwise. Putting it all together, our problem can be formulated as follows:
\begin{equation}
\setlength{\belowdisplayskip}{0pt} \setlength{\belowdisplayshortskip}{0pt}
\setlength{\abovedisplayskip}{0pt} \setlength{\abovedisplayshortskip}{0pt} 
\begin{aligned}
& \underset{\phi\in \Phi}{\text{minimize}}
& & \lim_{T\to+\infty} \text{sup}\:\frac{1}{T}\mathbb{E}^{\phi}\Big(\sum_{t=0}^{T-1}S^{\phi}(t)|S(0)\Big),\\
& \text{subject to}
& & \lim_{T\to+\infty} \text{sup}\:\frac{1}{T}\mathbb{E}^{\phi}\Big(\sum_{t=0}^{T-1}\psi^{\phi}(t)\Big)\leq \alpha.
\end{aligned}
\label{constrainedobjective}
\end{equation}
To address the above problem, we proceed with a Lagrange approach that transforms our constrained minimization problem into an optimization of the Lagrangian function. More specifically, by letting $\lambda\in\mathbb{R}^{+}$ be the Lagrange multiplier, we define the Lagrangian function as follows:
\begin{equation}
f(\lambda,\phi)=\lim_{T\to+\infty} \text{sup}\:\frac{1}{T}\mathbb{E}^{\phi}\Big(\sum_{t=0}^{T-1}S^{\phi}(t)+\lambda\psi^{\phi}(t)|S(0)\Big)-\lambda\alpha.
\label{lagrangeobjective}
\end{equation}
To that extent, the Lagrange approach can be summarized in the following problem:
\begin{equation}
 \underset{\lambda\in\mathbb{R}^{+}}{\text{max}}\:\: \underset{\phi\in \Phi}{\text{min}}\:\:f(\lambda,\phi).
\label{maxmin}
\end{equation}
It is well-known that for any feasible scheduling policy $\phi$ satisfying the constraint in (\ref{constrainedobjective}), the optimal value of the problem in (\ref{maxmin}) forms a lower bound to that of our original problem in (\ref{constrainedobjective}). The difference between the two values is known as the \emph{duality} gap, which is generally non-zero. Our goal is to show that our approach can achieve the optimal solution of the problem in (\ref{constrainedobjective}). To that extent, we first study in the sequel the problem:
\begin{equation}
g(\lambda)=\underset{\phi\in \Phi}{\text{min}}\:\:f(\lambda,\phi).
\label{justmin}
\end{equation}
\subsection{MDP Characterization}
Similarly to the previous section, we cast the problem (\ref{justmin}) into an MDP, which is the same as the one reported in the previous section except for the cost that is defined in this case as:
\begin{equation}
C(S(t),\psi(t))=S(t)+\lambda\psi(t).
\end{equation}
Following the same line of work, we know that the optimal policy $\phi^*$ of the problem $\underset{\phi\in \Phi}{\text{min}}\:\:f(\lambda,\phi)$ can be obtained by solving the Bellman equation for all $S\in\mathbb{N}$:
\begin{equation}
\theta_1 + V_1(S)=\min_{\psi\in\{0,1\}}\big\{S+\lambda\psi+\sum_{S'\in\mathbb{N} }\Pr(S\rightarrow S'|\psi)V_1(S')\big\},
\label{bellmannew}
\end{equation}
where $\Pr(S\rightarrow S'|\psi)$ is the transition probability from state $S$ to $S'$ given the action $\psi$, $\theta_1$ is the optimal value of the problem and $V_1(S)$ is the value function. As it was detailed in the previous section, solving the above equation directly is cumbersome in terms of complexity and hence, we provide structural properties of the optimal transmission policy in the next subsection.
\subsection{Structural Results}
In the same spirit as the previous section, we start by investigating the particularity of the value function $V_1(.)$.
\begin{lemma}
The value function $V_1(S)$ is increasing in $S$.
\label{increasingvaluelambda}
\end{lemma}
\begin{IEEEproof}
The proof follows the same procedure of Lemma \ref{increasingvalue} and is therefore omitted for the sake of space.
\end{IEEEproof}
The above lemma will be used to show that the optimal policy of our problem is a threshold policy. Before providing the proof of our claim, we first lay out the following definition.
\begin{definition}
An increasing threshold policy is a deterministic stationary policy in which the transmitter remains idle if the current state of the system $S$ is smaller than $n$ and attempts to transmit otherwise. In this case, the policy is fully characterized by the threshold $n\in\mathbb{N}$.
\end{definition}
With the above definition being laid out, we present the following proposition.
\begin{proposition}
The optimal policy $\phi^*$ of the problem in (\ref{justmin}) is an increasing threshold policy.
\label{propositionthreshold}
\end{proposition}
\begin{IEEEproof}
The proof can be found in Appendix \ref{proofproposition1} of the
supplementary material.
\end{IEEEproof}
With the structure of the optimal policy of (\ref{justmin}) being found, we tackle in more depth the average cost of our MDP when a threshold policy is adopted. To that extent, we recall that a threshold policy is fully characterized by its threshold value $n$. Accordingly, our problem in (\ref{justmin}) can be reformulated as follows:
\begin{equation}
\begin{aligned}
& \underset{n\in \mathbb{N}}{\text{minimize}} 
& & \overline{C}(n,\lambda),
\end{aligned}
\label{thresholdobjective}
\end{equation}
where $\overline{C}(n,\lambda)$ is the infinite horizon average cost of the MDP when the threshold policy is adopted. To find the expression of $\overline{C}(n,\lambda) \:\:\forall n\in\mathbb{N}$, we first tackle the special case where the transmitter always send updates at each time slot (i.e., $n=0$). In this scenario, the portion of time where the transmitter is sending updates, which is defined as $\lim_{T\to+\infty} \text{sup}\:\frac{1}{T}\mathbb{E}^{\phi}\Big(\sum_{t=0}^{T-1}\psi^{\phi}(t)\Big)$, is equal to $1$. Moreover, by using Theorem \ref{optimalpolicyunconstrained}, we end up with the following:
\begin{equation}
\overline{C}(0,\lambda)=(N-1)p_t\frac{\frac{1}{(1-a)^2}}{1+\frac{(N-1)p_t}{1-a}}+\lambda(1-\alpha),
\end{equation}
with $a$ being equal to $p_Rp_f+(N-2)p_t+p_sp_t$. Next, we shift our attention to the case where $n\in\mathbb{N}^*$. To that extent, we note that for any threshold policy, the MDP can be modeled through a Discrete Time Markov Chain (\textbf{DTMC}) where:
\begin{itemize}
\item The states refer to the values of the penalty function $S(t)$.
\item For any state $S(t)<n$, the transmitter is idle and therefore the dynamics of $S(t)$ coincide with those of $\psi(t)=0$ of Section III-B. On the other hand, for any state $S(t)\geq n$, the dynamics of $S(t)$ coincide with those of $\psi(t)=1$ of the same section.
\end{itemize}
Consequently, we focus in the sequel on this DTMC.
\begin{figure}[!ht]
\centering
\includegraphics[width=.9\linewidth]{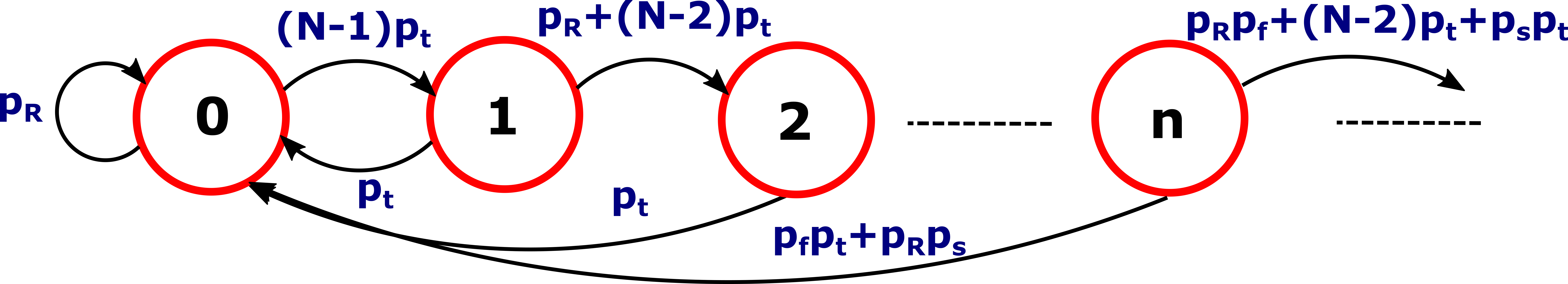}
\setlength{\belowcaptionskip}{-5pt}
\caption{The states transitions under a threshold policy.}
\label{thresholddtmc}
\end{figure}\\
The next step towards finding the average cost $\overline{C}(n,\lambda)$ consists of calculating the stationary distribution of the DTMC. We, therefore, provide the following proposition.
\begin{proposition}
For a fixed threshold $n\in\mathbb{N}^{*}$, the DTMC in question is irreducible and admits $\pi_k(n) \:\: \forall k\in\mathbb{N}$ as its stationary distribution where:
\begin{equation}
\pi_0(n)=\frac{1}{1+\frac{(N-1)p_t(1-b^n)}{1-b}+\frac{(N-1)p_tab^{n-1}}{1-a}},
\label{pi0}
\end{equation}
\begin{equation}
\pi_k(n)=(N-1)p_tb^{k-1}\pi_0 \quad 1\leq k\leq n,
\label{pik}
\end{equation}
\begin{equation}
\pi_k(n)=(N-1)p_tb^{n-1}a^{k-n}\pi_0 \quad k\geq n+1,
\label{pin}
\end{equation}
with $a,b$ being two constants that are equal to $p_Rp_f+(N-2)p_t+p_sp_t$ and $p_R+(N-2)p_t$ respectively.
\label{stationarydistribution}
\end{proposition}
\begin{IEEEproof}
The proof can be found in Appendix \ref{proofproposition2} of the
supplementary material.
\end{IEEEproof}
By leveraging Proposition \ref{stationarydistribution}, we can proceed to find a closed form of the average cost of the threshold policy.
\begin{theorem}
For a fixed threshold $n\in\mathbb{N}^{*}$, the average cost of the policy is $\overline{C}(n,\lambda)=\overline{C}(n)+\overline{C}_1(n,\lambda)$ where:
\begin{equation}
\overline{C}(n)=(N-1)p_t\frac{\frac{1+b^n(nb-n-1)}{(1-b)^2}+\frac{b^{n-1}a(n+\frac{1}{1-a})}{1-a}}{1+\frac{(N-1)p_t(1-b^n)}{1-b}+\frac{(N-1)p_tab^{n-1}}{1-a}},
\label{costgamma}
\end{equation}
\begin{equation}
\overline{C}_1(n,\lambda)=\lambda\frac{(N-1)p_tb^{n-1}}{(1-a)(1+\frac{(N-1)p_t(1-b^n)}{1-b}+\frac{(N-1)p_tab^{n-1}}{1-a})}
-\lambda\alpha.
\label{costlambda}
\end{equation}
\label{theoremcostgamma}
\end{theorem}
\begin{IEEEproof}
The proof can be found in Appendix \ref{prooftheorem2} of the
supplementary material.
\end{IEEEproof}
As we now have the expression of the average cost $\overline{C}(n,\lambda)$, we turn our attention to studying its characteristics in order to prove the optimality of the Lagrange approach.
\subsection{Optimality of the Lagrange Approach}
The optimality of the Lagrange approach in similar resource-constrained environments has been established in the literature for other cost functions (e.g., the AoI in \cite{8377368}). However, contrary to \cite{8377368}, the standard approach to prove this optimality cannot be adopted in our case. This is mainly due to the complexity of the average cost function reported in Theorem \ref{theoremcostgamma}. In particular, as seen in (\ref{costgamma})-(\ref{costlambda}), $\overline{C}(n,\lambda)$ is not necessarily convex in $n$, which limits the applicability of the approach adopted in \cite{8377368}. Accordingly, to demonstrate the optimality of the Lagrange approach in our case, we proceed in a different direction. Specifically, we investigate in more depth the behavior of the cost function and leverage these results to establish the AoII-optimal policy. To present our approach, we first let $\big(A(n)\big)_{n\in\mathbb{N}}$ be the portion of time where the transmitter is attempting to send packets. To that extent, we have that $\big(A(n)\big)_{n\in\mathbb{N}}$ is a decreasing positive sequence with $A(0)=1$ and $\big(A(n)\big)_{n\in\mathbb{N}^*}=\sum\limits_{k=n}^{+\infty}\pi_k(n)$ which can be expressed as:
\begin{equation}
A(n)=\frac{(N-1)p_tb^{n-1}}{(1-a)(1+\frac{(N-1)p_t(1-b^n)}{1-b}+\frac{(N-1)p_tab^{n-1}}{1-a})} \quad \forall n\in\mathbb{N}^{*}
\end{equation}
To that end, we have $\overline{C}_1(n,\lambda)=\lambda A(n) -\lambda\alpha$. With this definition in mind, we summarize our approach in the following:
\begin{enumerate}
\item We prove that $\overline{C}(n)$, which is reported in (\ref{costgamma}), is increasing with $n$. 
\item We define the set of intersection points 
\begin{equation}
\lambda(n)=\frac{\overline{C}(n+1)-\overline{C}(n)}{A(n)-A(n+1)} \quad \forall n\in\mathbb{N}
\end{equation}
\item We prove that $\lambda(n)$ is increasing with $n$.
\item We relate through graphical methods and several inductive lemmas the results on $\lambda(n)$ to the establishment of the AoII-optimal policy. 
\item We propose a low complexity algorithm to find the AoII-optimal operating point of the system.
\end{enumerate}
The details of the above steps will be laid out in the remainder of this section. To proceed in this direction, we also define $n(\lambda)$ as the optimum threshold that solves, for a fixed $\lambda$, the optimization problem in (\ref{thresholdobjective}). With the definitions dealt with, and with our steps being clarified, we now proceed with the proof of optimality. To that extent, let us first note that the following always holds: \color{black} 
\begin{equation}
g(\lambda)\leq \underset{\lambda\in \mathbb{R}^{+}}{\text{max}}\:\:g(\lambda)\leq \theta^*
\end{equation}
where $\theta^*$ is the optimal value of our constrained problem in (\ref{constrainedobjective}). Consequently, if we can find $\lambda_1$ such that $A(n(\lambda_1))=\alpha$, then $g(\lambda_1)=\underset{\lambda\in \mathbb{R}^{+}}{\text{max}}\:\:g(\lambda)=\theta^*$. In this case, we achieve the optimal operating point of (\ref{constrainedobjective}) by simply adopting a threshold policy characterized by the threshold $n(\lambda_1)$. However, the issue arises when such a value of $\lambda_1$ does not exist since the set $\{n(\lambda):\:\lambda\in\mathbb{R}^+\}$ is discrete. To deal with this case, we aim to show that we can always find $(n_0,\lambda_{n_0})$ such that: 
\begin{enumerate}
\item $\overline{C}(n_0,\lambda_{n_0})=\overline{C}(n_0+1,\lambda_{n_0})$
\item $\begin{cases}
 A(n_0)\geq\alpha\\
 A(n_0+1)<\alpha
\end{cases} $
\item $n(\lambda_{n_0})=n_0$
\end{enumerate}
In this case, it is sufficient to take a mixture of two threshold policies $\phi_{n_0}$ and $\phi_{n_0+1}$ with a probability $\rho=\frac{\alpha-A(n_0+1)}{A(n_0)-A(n_0+1)}$ and $1-\rho=\frac{A(n_0)-\alpha}{A(n_0)-A(n_0+1)}$ respectively, to achieve the optimal objective value of the constrained problem in (\ref{constrainedobjective}). We now proceed to show the existence and uniqueness of $(n_0,\lambda_{n_0})$.
\begin{proposition} The following always holds:
\begin{equation}
\forall n\in\mathbb{N}, \exists \lambda_{n}\in\mathbb{R}^{+}:  \overline{C}(n,\lambda_{n})=\overline{C}(n+1,\lambda_{n}).
\label{deflambda}
\end{equation}
\label{prooflambda}
\end{proposition}
\begin{IEEEproof} 
The proof can be found in Appendix \ref{proofproposition3}.
\end{IEEEproof}
As the above proposition holds for any $n$, let us focus on the value $n_0$ such that: 
\begin{equation}
\begin{cases}
 A(n_0)\geq\alpha\\
 A(n_0+1)<\alpha
\end{cases} 
\end{equation}
In the next theorem, we show that this value $n_0$ verifies $n(\lambda_{n_0})=n_0$.
\begin{theorem} For the aforementioned $\lambda_{n_0}$, $n_0$ minimizes the average cost function $\overline{C}(n,\lambda_{n_0})$.
\label{finalresultslambda}
\end{theorem}
\begin{IEEEproof}
The proof can be found in Appendix \ref{prooftheorem3}.
\end{IEEEproof}

\subsection{Algorithm Implementation}
Based on the previous section, we can assert that the optimal transmission policy consists of a mixture of two deterministic threshold policies $\phi_{n_0}$ and $\phi_{n_0+1}$ such that: 
\begin{equation}
\begin{cases}
 A(n_0)\geq\alpha\\
 A(n_0+1)<\alpha
\end{cases} 
\end{equation}
As $\big(A(n)\big)_{n\in\mathbb{N}}$ is a decreasing sequence in $n$, we can rewrite $n'=n_0+1$ as follows:
\begin{equation}
n'=\inf\{n\geq1: A(n)-\alpha<0\}.
\end{equation}
For any $0<\alpha\leq1$, we can attest that there exists a finite $n'$ that verifies the above condition. To find this value, we employ a two steps algorithm depicted in Algorithm $1$. The two steps are as follows:
\begin{itemize}
\item Exponential increase of the upperbound value $N_{UB}$ to ensure that $n'$ is included in the interval of interest $[N_{LB},N_{UB}]$.
\item A binary search in the interval mentioned above to find the value $n'$.
\end{itemize}
\begin{algorithm}
\caption{Optimal threshold finder}\label{euclid}
\begin{algorithmic}[1]
\Procedure{Upperbound increase}{}
\State \textbf{Init.} $N_{LB}=N_{UB}=1$
\While {$A(N_{UB})-\alpha\geq0$}
\State $N_{LB}:=N_{UB}$
\State $N_{UB}:=2N_{UB}$
\EndWhile
\EndProcedure
\Procedure{Binary search}{}
\State $n':=\ceil*{\frac{N_{LB}+N_{UB}}{2}} $
\While {$n'<N_{UB}$}
\If {$A(n')-\alpha\geq0$} $N_{LB}:=n'$
\Else $\:\:N_{UB}=n'$
\EndIf
\State $n':=\ceil*{\frac{N_{LB}+N_{UB}}{2}} $
\EndWhile
\EndProcedure
\State Output the optimal threshold $n_0=n'-1$
\end{algorithmic}
\end{algorithm}
\setlength{\textfloatsep}{0pt}
The first part of the algorithm finishes in $N_1=\log_{2}(n')$ iterations while the binary search part is known to have a worst-case complexity of $\log_2(N_{size})$ where $N_{size}$ is the size of the interval of interest. To that extent, we have that: $N_{size}=2^{N_1}-2^{N_1-1}$. Hence, the worst-case complexity of the second part is $\log_2(N_{size})=N_1-1$. We can, therefore, conclude that the complexity of the above algorithm is logarithmic in the value of $n'$, which makes it appealing to be implemented in practice.

After the algorithm finishes and $n_0$ is found, it is sufficient to adopt a transmission policy where a packet is generated and transmitted when the penalty is equal to $n_0$ and $n_0+1$ with a probability $\rho=\frac{\alpha-A(n_0+1)}{A(n_0)-A(n_0+1)}$ and $1-\rho=\frac{A(n_0)-\alpha}{A(n_0)-A(n_0+1)}$ respectively to achieve the optimal objective value of the constrained problem in (\ref{constrainedobjective}).
\section{Numerical Results}
In this section, we provide numerical results that highlight the effects of the information source dynamics on the performance of our proposed AoII-optimal policy. We also compare our framework to both the AoI and the error function minimization frameworks in order to shed light on important insights. Note that, although we focus on the Markovian information source depicted in Section III-A, the insights provided in this section intuitively hold for more general information source models.
\color{black}
\subsection{Information Source Parameters}
In the first scenario, we investigate in more depth the effect of the Markov chain's dynamics on the performance of our proposed AoII-optimal policy. 
\subsubsection{Effect of $p_R$}
\begin{figure}[!ht]
\centering
\includegraphics[width=.65\linewidth]{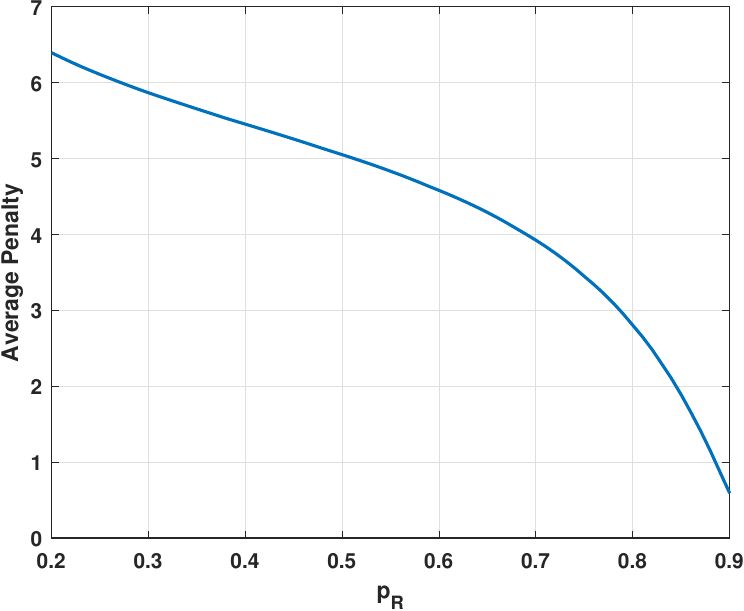}
\setlength{\belowcaptionskip}{-5pt}
\caption{The average AoII in function of $p_R$.}
\label{simulationscenarioPr}
\end{figure}

In this scenario, we consider that the number of states is $N=8$, and we fix the parameter $\alpha$ to $0.1$. As for the channel parameter, we assume that the transmission success probability $p_s$ is equal to $0.8$. While making sure that $p_t<p_R$, we vary the probability of remaining in the same state $p_R$ and plot the average AoII of the optimal policy. As seen in Fig. \ref{simulationscenarioPr}, the average cost decreases as $p_R$ increases. The reason behind this is twofold:
\begin{enumerate}
\item When $p_R$ is high, the information source becomes more ``predictable". In other words, when a packet is transmitted, it is less likely for it to become obsolete due to a transition of the Markov chain during the transmission stage.
\item When $p_R$ is high, the AoII remains zero for a significant amount of time upon successful transmission. This allows us to make better use of the permitted power budget $\alpha$ as we will be able to transmit at a lower threshold value without exceeding the allowed power budget. This can be verified by looking at $n_0$ in function of $p_R$ in the following table:
\begin{center}
\begin{tabular}{|c|c|c|c|c|}
 \hline
  $p_R$ & $0.2$  & $0.4$  & $0.6$  & $0.8$ \\
  \hline
 $n_0$ & $15$  & $12$  & $10$  & $7$ \\
 \hline
\end{tabular}
 \captionsetup{justification=centering}
\captionof{table}{Variation of $n_0$ in function of $p_R$.}
 \label{RRCSMA}
\end{center}
We can see from the above table that as $p_R$ increases, the value of $n_0$ decreases. In other words, our tolerance for the value of the AoII is reduced, and we can transmit at a much lower AoII value without violating the power constraint. This eventually leads to a reduction in the average AoII.
\end{enumerate}
\subsubsection{Effect of $N$}
\begin{figure}[!ht]
\centering
\includegraphics[width=.6\linewidth]{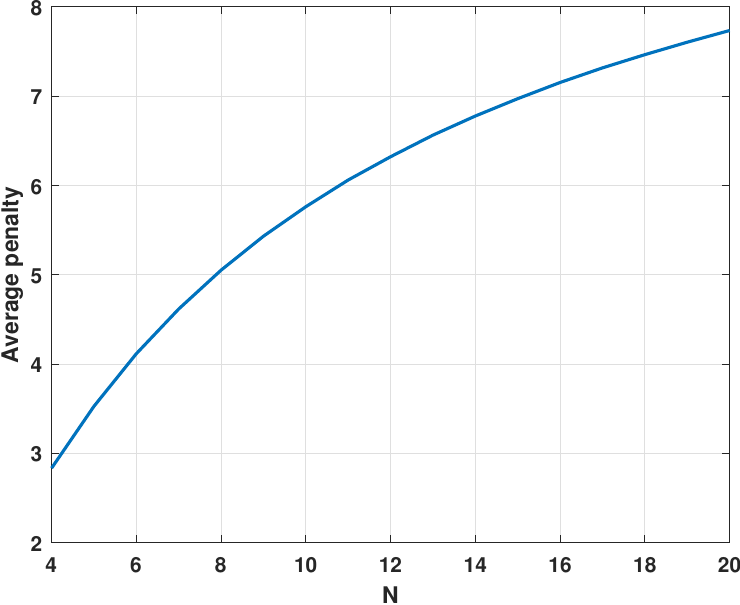}
\caption{The average AoII in function of $N$.}
\setlength{\belowcaptionskip}{-5pt}
\label{infunctionn}
\end{figure}

We consider the case where $p_R=0.5$, $\alpha=0.1$, and the probability of successful transmission is $p_s=0.8$.  We vary $N$ and report the average AoII when the AoII-optimal policy is adopted in Fig. \ref{infunctionn}. As can be seen in the figure, the average AoII increases when the number of states $N$ grows. To explain this trend, we first recall that the transition probabilities at each state always verify the following equality:
\begin{equation}
p_R+(N-1)p_t=1
\label{formulaaa}
\end{equation}
Accordingly, we can use (\ref{formulaaa}) to conclude that $p_t=\frac{1-p_R}{N-1}$. Next, let us consider that the monitor has perfect knowledge of the information process at time $t$, denoted by $X(t)$. Then, let us suppose that the information source changes value at time $t+1$, which happens with a fixed probability $1-p_R$. With that in mind, we recall that the probability for the information source to go back to its old value $X(t)$ at time $t+2$ is $p_t$. As $p_t$ is a decreasing function in $N$, this means that the probability for the monitor to have correct knowledge of the information source at time $t+2$ without wasting resources for packet transmission, decreases with $N$. Accordingly, when $N$ grows, the average AoII will also increase.
\color{black}
\subsection{Comparison with the AoI Framework}
In the following, we provide a comparison between our optimal transmission policy and the optimal age policy of \cite{8377368}. 
\subsubsection{Comparison in Function of $\alpha$}
\begin{figure}[!ht]
\centering
\includegraphics[width=.65\linewidth]{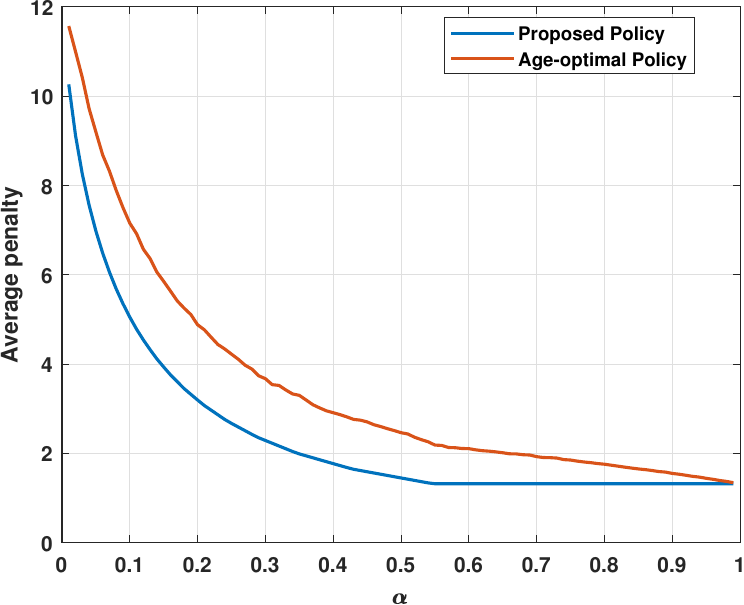}
\setlength{\belowcaptionskip}{-5pt}
\caption{Comparison between our proposed policy and the age-optimal transmission policy in terms of average AoII.}
\label{simulationscenario1}
\end{figure}

We adopt in this case the same number of states $N=8$ and success probability $p_s=0.8$. We fix the probability of remaining in the same state $p_R$ to $0.5$. We vary the parameter $\alpha$ and plot the average AoII achieved by both policies. As seen in Fig. \ref{simulationscenario1}, the proposed policy always outperforms the age-optimal policy for all values of $\alpha$. The following two observations can also be drawn from the figure:
\begin{enumerate}
\item One can see that the two curves converge as $\alpha$ increases. This is in agreement with our theoretical results in the unconstrained case in Section IV. In fact, when the imposed power constraint becomes less restrictive, the transmitter will be sending more
packets and we converge to the “always update” policy that
minimizes both the AoII and the AoI.
\item Another interesting observation is that the gap between the two curves is small when $\alpha$ is small (e.g., the gap is equal to $1.1$ for $\alpha=0.02$). This is due to the number of packets sent by the transmitter becoming very small. Consequently, the average AoII will be mostly dictated by how the Markov chain evolves rather than the transmission policy adopted. Therefore, in this case, we converge to the ``no updates" average cost previously reported in eq. (\ref{noupdatescost}).
\end{enumerate}
By combining the above two observations, we can conclude that when the transmitter is heavily constrained by its power, or when it has unlimited power, age-optimal policies lead to virtually the same performance as the optimal AoII policy.

\begin{figure}[!ht]
\centering
\includegraphics[width=.65\linewidth]{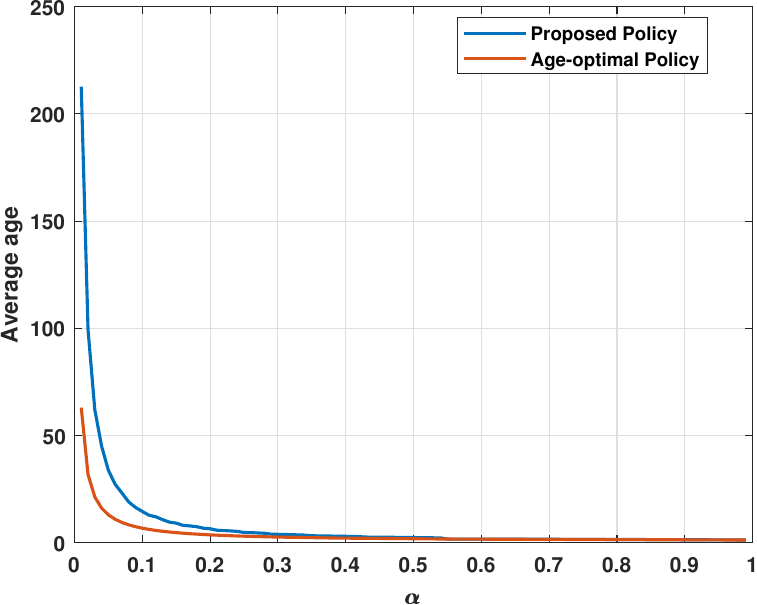}
\setlength{\belowcaptionskip}{-10pt}
\caption{Comparison between the two policies in terms of average age.}
\label{simulationscenario2}
\end{figure}

We also investigate the age performance of our proposed policy and compare it to the age-optimal policy. As seen in Fig. \ref{simulationscenario2}, the age-optimal policy outperforms our policy in terms of average age. However, the gap between the two curves vanishes for high $\alpha$ and that is for the same reason previously reported in the average AoII comparison between the two policies. On the other hand, as $\alpha$ decreases, the gap between the two curves increases, reaching $190$ for $\alpha=0.02$. The reason behind this is the fact that as $\alpha$ decreases, the allowed number of transmissions becomes extremely small. Therefore, the impact of the transmission decisions will become more significant on the performance. To that extent, since our policy is based on the information content of the packet rather than just the age at the monitor, our proposed penalty measure can sometimes be equal to $0$ while the age is equal to $100$. The differences of spirit between the two transmission polices will lead to a significant difference in age performance when the available power budget is really small.
\setlength{\textfloatsep}{0pt}
\subsubsection{Comparison in Function of $p_R$}
\begin{figure}[!ht]
\centering
\includegraphics[width=.6\linewidth]{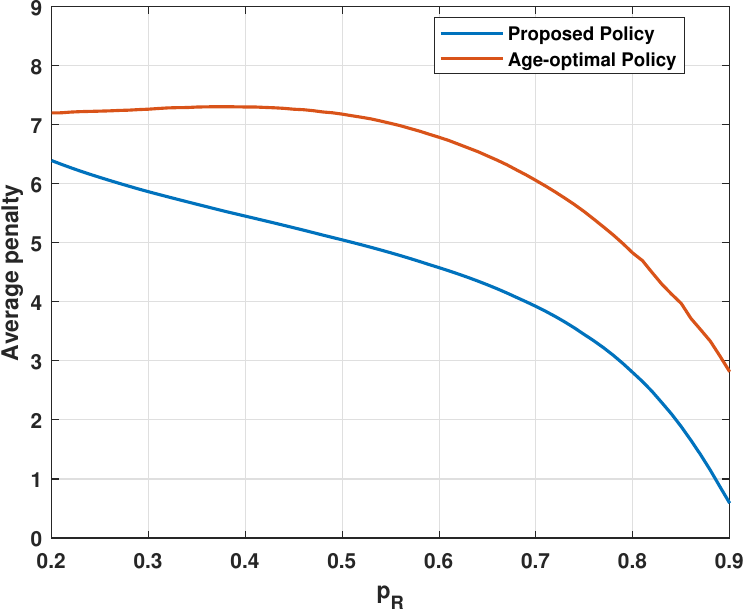}
\caption{Comparison in function of $p_R$ between the AoII-optimal policy and the AoI-optimal policy.}
\setlength{\belowcaptionskip}{-5pt}
\label{infunctionpr}
\end{figure}

In this scenario, we compare the AoII-optimal policy and the AoI-optimal policy when $p_R$ is varied. We consider that $N=8$, $\alpha=0.1$, and the probability of successful transmission is $p_s=0.8$. While maintaining $p_t<p_R$, we vary $p_R$ and report the differences between the two policies in Fig. \ref{infunctionpr}. As can be seen in the figure, the gap between the two curves increases as $p_R$ grows (from $0.7$ for $p_R=0.2$ to $2.2$ for $p_R=0.9$). To explain this, we recall that the AoI always increases regardless of the value of the information source. As $p_R$ increases, the information process $X(t)$ will have a higher probability of keeping the same value at the next time slot $t+1$. However, since the AoI is always increasing, the AoI-optimal policy will waste vital resources to update the monitor when it is not necessary to do so. As the AoI-optimal policy sends more obsolete packets when $p_R$ is high, this will to a non-negligible gap between the AoI-optimal and AoII-optimal policies as seen in the figure.
\subsection{Comparison with the Error Framework}
We present in the following a comparison between our policy and the error-based policy that follows the rules below:
\begin{itemize}
\item Send a packet solely when the monitor has a wrong estimate of the information source.
\item Ensure that the constraint on the power consumption is verified. with \emph{equality}.
\end{itemize}
We consider the case where $N=8$, $p_R=0.5$, and the probability of successful transmission is $p_s=0.8$. We report in the next table the AoII values of the two policies. As can be seen in Table \ref{RRCSMAasda}, our policy always outperforms the error based policy. We can also see that as $\alpha$ increases, the gap between the two shrinks since the transmitter will be sending more packets and we converge to the ``always update" policy that minimizes both the AoII and the status error function. 
\begin{center}
\begin{tabular}{|c|c|c|}
 \hline
 $\alpha$ & $\text{AoII}_{\text{optimal}}$ & $\text{AoII}_{\text{error}}$ \\
  \hline
 $0.12$  & $4.4$ & $5.9$ \\
 $0.25$  & $2.7$ & $3.8$\\
 $0.45$  & $2$ & $2.2$\\
 \hline
\end{tabular}
 \captionof{table}{Comparison between the AoII-optimal policy and the error based policy.}
 \label{RRCSMAasda}
\end{center}
\color{black}
\vspace{-5pt}
\section{Conclusion and Future Work}
In this paper, we have proposed a new performance metric that deals with the shortcomings of the conventional AoI and error penalty functions in the framework of status updates. Dubbed as the Age of Incorrect Information, this new metric extends the notion of fresh updates and adequately captures the information content that the updates bring to the monitor. We have studied the metric mentioned above in the case where a transmitter-receiver pair communicates over an unreliable channel. By leveraging MDP tools, the optimal policy's structure was found for the cases where the transmitter is limited and non-limited by its power. A low complexity algorithm was then presented that finds the optimal operating point that minimizes the average AoII. Lastly, numerical results were laid out that highlight the effect of the information source's dynamics on the AoII, along with a comparison between the AoI and AoII frameworks. The analysis in this paper can be used as a basis for the multi-user case, similar to the work done in \cite{2020arXiv200103096M}. Other future research directions include the extension to more general information source models, the investigation of broader choices of time-dissatisfaction and error functions, and the examination of continuous-time systems. \color{black} 
\bibliographystyle{IEEEtran}
\bibliography{trialout}

\begin{thebibliography}{10}
\providecommand{\url}[1]{#1}
\csname url@samestyle\endcsname
\providecommand{\newblock}{\relax}
\providecommand{\bibinfo}[2]{#2}
\providecommand{\BIBentrySTDinterwordspacing}{\spaceskip=0pt\relax}
\providecommand{\BIBentryALTinterwordstretchfactor}{4}
\providecommand{\BIBentryALTinterwordspacing}{\spaceskip=\fontdimen2\font plus
\BIBentryALTinterwordstretchfactor\fontdimen3\font minus
  \fontdimen4\font\relax}
\providecommand{\BIBforeignlanguage}[2]{{%
\expandafter\ifx\csname l@#1\endcsname\relax
\typeout{** WARNING: IEEEtran.bst: No hyphenation pattern has been}%
\typeout{** loaded for the language `#1'. Using the pattern for}%
\typeout{** the default language instead.}%
\else
\language=\csname l@#1\endcsname
\fi
#2}}
\providecommand{\BIBdecl}{\relax}
\BIBdecl

\bibitem{5307471}
P.~Papadimitratos, A.~D.~L. Fortelle, K.~Evenssen, R.~Brignolo, and S.~Cosenza,
  ``Vehicular communication systems: Enabling technologies, applications, and
  future outlook on intelligent transportation,'' \emph{IEEE Communications
  Magazine}, vol.~47, no.~11, pp. 84--95, November 2009.

\bibitem{5597912}
P.~Corke, T.~Wark, R.~Jurdak, W.~Hu, P.~Valencia, and D.~Moore, ``Environmental
  wireless sensor networks,'' \emph{Proceedings of the IEEE}, vol.~98, no.~11,
  pp. 1903--1917, Nov 2010.

\bibitem{6195689}
S.~Kaul, R.~Yates, and M.~Gruteser, ``Real-time status: How often should one
  update?'' in \emph{2012 Proceedings IEEE INFOCOM - IEEE Conference on
  Computer Communications}, March 2012, pp. 2731--2735.

\bibitem{2018arXiv180202129A}
A.~{Arafa}, J.~{Yang}, S.~{Ulukus}, and H.~V. {Poor}, ``{Age-Minimal Online
  Policies for Energy Harvesting Sensors with Incremental Battery Recharges},''
  \emph{ArXiv e-prints}, p. arXiv:1802.02129, Feb. 2018.

\bibitem{8000687}
Y.~Sun, E.~Uysal-Biyikoglu, R.~D. Yates, C.~E. Koksal, and N.~B. Shroff,
  ``Update or wait: How to keep your data fresh,'' \emph{IEEE Transactions on
  Information Theory}, vol.~63, no.~11, pp. 7492--7508, Nov 2017.

\bibitem{8006506}
R.~D. {Yates}, P.~{Ciblat}, A.~{Yener}, and M.~{Wigger}, ``Age-optimal
  constrained cache updating,'' in \emph{2017 IEEE International Symposium on
  Information Theory (ISIT)}, June 2017, pp. 141--145.

\bibitem{8006590}
Y.~Hsu, E.~Modiano, and L.~Duan, ``Age of information: Design and analysis of
  optimal scheduling algorithms,'' in \emph{2017 IEEE International Symposium
  on Information Theory (ISIT)}, June 2017, pp. 561--565.

\bibitem{2018arXiv180704356Z}
B.~{Zhou} and W.~{Saad}, ``{Joint Status Sampling and Updating for Minimizing
  Age of Information in the Internet of Things},'' \emph{arXiv e-prints}, p.
  arXiv:1807.04356, Jul 2018.

\bibitem{8845254}
A.~{Maatouk}, M.~{Assaad}, and A.~{Ephremides}, ``Minimizing the age of
  information: Noma or oma?'' in \emph{IEEE INFOCOM 2019 - IEEE Conference on
  Computer Communications Workshops (INFOCOM WKSHPS)}, April 2019, pp.
  102--108.

\bibitem{2020arXiv200103096M}
A.~{Maatouk}, S.~{Kriouile}, M.~{Assaad}, and A.~{Ephremides}, ``{On The
  Optimality of The Whittle's Index Policy For Minimizing The Age of
  Information},'' \emph{arXiv e-prints}, p. arXiv:2001.03096, Jan. 2020.

\bibitem{8377368}
E.~T. {Ceran}, D.~{Gündüz}, and A.~{György}, ``Average age of information
  with hybrid arq under a resource constraint,'' in \emph{2018 IEEE Wireless
  Communications and Networking Conference (WCNC)}, April 2018, pp. 1--6.

\bibitem{2018arXiv180306469T}
R.~{Talak}, S.~{Karaman}, and E.~{Modiano}, ``{Distributed Scheduling
  Algorithms for Optimizing Information Freshness in Wireless Networks},''
  \emph{arXiv e-prints}, p. arXiv:1803.06469, Mar. 2018.

\bibitem{2018arXiv180103975J}
Z.~{Jiang}, B.~{Krishnamachari}, X.~{Zheng}, S.~{Zhou}, and Z.~{Niu}, ``{Timely
  Status Update in Massive IoT Systems: Decentralized Scheduling for Wireless
  Uplinks},'' \emph{arXiv e-prints}, p. arXiv:1801.03975, Jan. 2018.

\bibitem{2019arXiv190100481M}
A.~{Maatouk}, M.~{Assaad}, and A.~{Ephremides}, ``{Minimizing The Age of
  Information in a CSMA Environment},'' \emph{arXiv e-prints}, p.
  arXiv:1901.00481, Jan 2019, to appear in the 14th International Symposium on
  Modeling and Optimization in Mobile, Ad Hoc, and Wireless Networks (WiOpt
  2019).

\bibitem{9007478}
------, ``On the age of information in a csma environment,'' \emph{IEEE/ACM
  Transactions on Networking}, pp. 1--14, 2020.

\bibitem{2018arXiv180805738Z}
J.~{Zhong}, R.~D. {Yates}, and E.~{Soljanin}, ``{Multicast With Prioritized
  Delivery: How Fresh is Your Data?}'' \emph{ArXiv e-prints}, p.
  arXiv:1808.05738, Aug. 2018.

\bibitem{2018arXiv180104067N}
E.~{Najm}, R.~{Nasser}, and E.~{Telatar}, ``{Content Based Status Updates},''
  \emph{ArXiv e-prints}, p. arXiv:1801.04067, Jan. 2018.

\bibitem{8613408}
A.~{Maatouk}, M.~{Assaad}, and A.~{Ephremides}, ``The age of updates in a
  simple relay network,'' in \emph{2018 IEEE Information Theory Workshop
  (ITW)}, Nov 2018, pp. 1--5.

\bibitem{8849695}
------, ``Age of information with prioritized streams: When to buffer preempted
  packets?'' in \emph{2019 IEEE International Symposium on Information Theory
  (ISIT)}, July 2019, pp. 325--329.

\bibitem{2020arXiv200201916M}
A.~{Maatouk}, Y.~{Sun}, A.~{Ephremides}, and M.~{Assaad}, ``{Status Updates
  with Priorities: Lexicographic Optimality},'' \emph{arXiv e-prints}, p.
  arXiv:2002.01916, Feb. 2020.

\bibitem{8006542}
Y.~{Sun}, Y.~{Polyanskiy}, and E.~{Uysal-Biyikoglu}, ``Remote estimation of the
  wiener process over a channel with random delay,'' in \emph{2017 IEEE
  International Symposium on Information Theory (ISIT)}, June 2017, pp.
  321--325.

\bibitem{2018arXiv181205215J}
Z.~{Jiang}, S.~{Zhou}, Z.~{Niu}, and Y.~{Cheng}, ``{A Unified Sampling and
  Scheduling Approach for Status Update in Multiaccess Wireless Networks},''
  \emph{arXiv e-prints}, p. arXiv:1812.05215, Dec 2018.

\bibitem{8437927}
J.~{Zhong}, R.~D. {Yates}, and E.~{Soljanin}, ``Two freshness metrics for local
  cache refresh,'' in \emph{2018 IEEE International Symposium on Information
  Theory (ISIT)}, June 2018, pp. 1924--1928.

\bibitem{8406891}
C.~{Kam}, S.~{Kompella}, G.~D. {Nguyen}, J.~E. {Wieselthier}, and
  A.~{Ephremides}, ``Towards an effective age of information: Remote estimation
  of a markov source,'' in \emph{IEEE INFOCOM 2018 - IEEE Conference on
  Computer Communications Workshops (INFOCOM WKSHPS)}, April 2018, pp.
  367--372.

\bibitem{2018arXiv180101803K}
I.~{Kadota}, A.~{Sinha}, E.~{Uysal-Biyikoglu}, R.~{Singh}, and E.~{Modiano},
  ``{Scheduling Policies for Minimizing Age of Information in Broadcast
  Wireless Networks},'' \emph{arXiv e-prints}, p. arXiv:1801.01803, Jan 2018.

\bibitem{Bertsekas:2000:DPO:517430}
D.~P. Bertsekas, \emph{Dynamic Programming and Optimal Control}, 2nd~ed.\hskip
  1em plus 0.5em minus 0.4em\relax Athena Scientific, 2000.

\end{thebibliography}
\appendices
\section{Proof of Proposition \ref{prooflambda}}
\label{proofproposition3}
\begin{figure}[!ht]
\centering
\includegraphics[width=.73\linewidth]{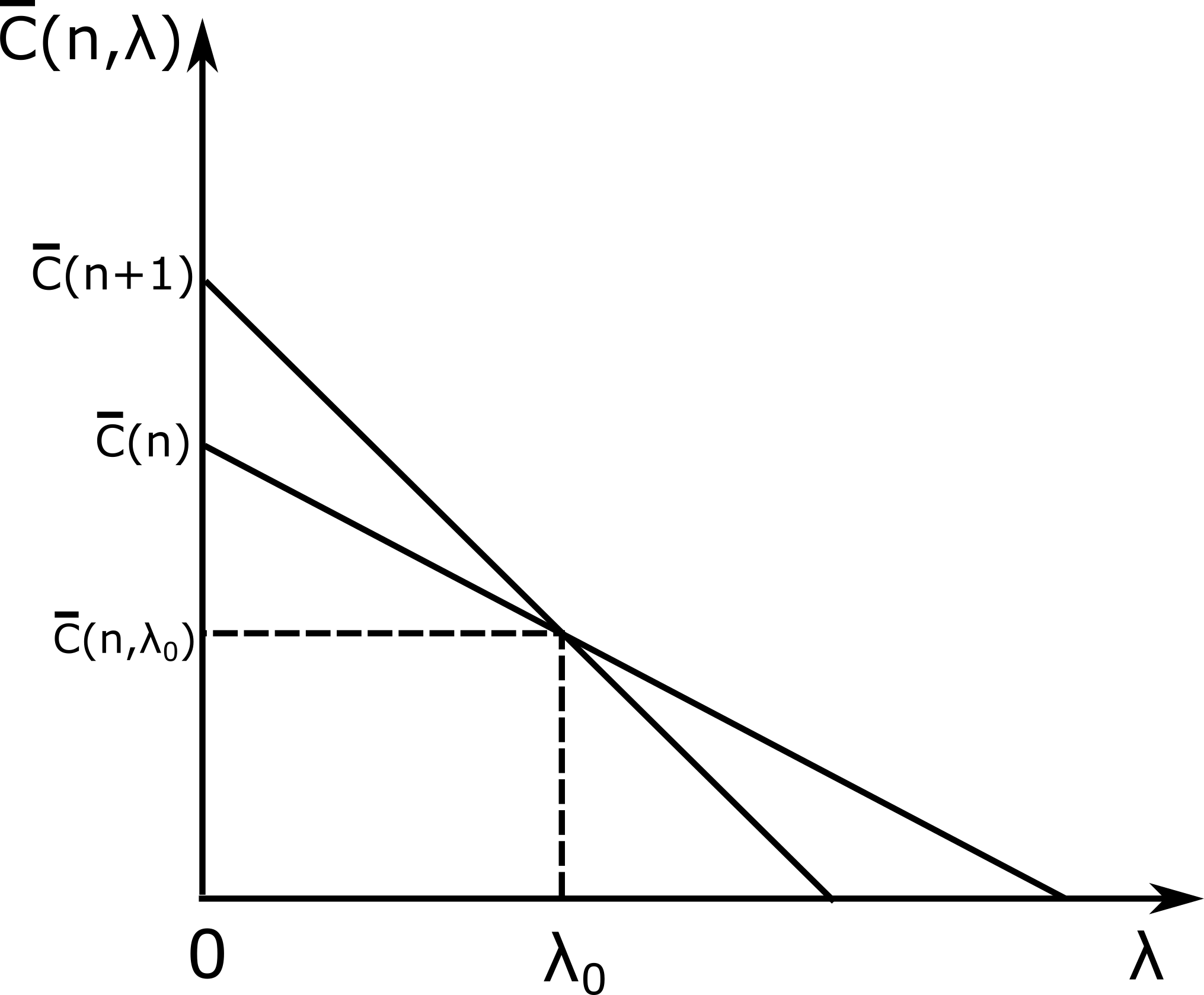}
\setlength{\belowcaptionskip}{-5pt}
\caption{Illustration of the intersection proof.}
\label{prooflambdaillu}
\end{figure}

Before investigating the general scenario, we first note that the proposition is trivially true for $n=0$. In fact, we first note that $\overline{C}(n)$ is nothing but the average penalty of a threshold policy in the unconstrained MDP case reported in Section IV-B. As $\overline{C}(0)=\overline{C}(1)$ (we refer the readers to the results of Theorem \ref{optimalpolicyunconstrained}), we can easily verify that we have $\overline{C}(0,0)=\overline{C}(1,0)$. To tackle the case where $n\in\mathbb{N}^*$, we provide a proof that revolves around a graphical illustration in Fig. \ref{prooflambdaillu} of $\overline{C}(n,\lambda)$ in function of $\lambda$. To proceed in that direction, we first study in the next lemma the variation of $\overline{C}(n)$ in function of $n,\forall n\in\mathbb{N}^*$.
\begin{lemma}
The function $\overline{C}(n)$ is increasing with $n$.
\label{lemmacnincreasing}
\end{lemma}
\begin{IEEEproof}
By considering the expression of $\overline{C}(n)$ previously reported in (\ref{costgamma}), we can observe that it is rather difficult to study its variations directly. To circumvent this difficulty, we recall that $\overline{C}(n)$ is nothing but the average penalty of a threshold policy in the unconstrained MDP case reported in Section IV-B. The dynamics of such a threshold policy is identical to the DTMC reported in Fig. \ref{thresholddtmc}. By observing the DTMC in question, we can see that the chain can only move backward due to a transition to state $0$. When the transmitter does not attempt to send a packet ($S<n$), the probability of transition to state $0$ is $p_t$. However, when the transmitter sends packets ($S\geq n$), the probability of reducing the penalty to zero is $p_fp_t+p_Rp_s$. As $p_R>p_t$, we can conclude that $p_fp_t+p_Rp_s>p_t$. Consequently, a transmission of a packet will always increase the likelihood of transitions to the state $0$. Based on this, we can conclude that employing a higher threshold, which leads to a smaller number of transmissions, will undoubtedly increase the average penalty. 
\end{IEEEproof}
By using the above results, and as $\overline{C}(n,0)=\overline{C}(n)$, we can conclude that the points on the $y-$axis in Fig. \ref{prooflambdaillu} move upwards as $n$ increases. Moreover, by using the expression of $\overline{C}(n,\lambda)$ in Theorem \ref{theoremcostgamma}, we can deduce that the slope of $\overline{C}(n,\lambda)$ is nothing but $A(n)-\alpha$. Since $A(n)$ decreases when the threshold $n$ increases, we can assert that the slope of the curves $\overline{C}(n,\lambda)$ decreases with $n$. By combining the above two observations, we can see that for any fixed value of $n$, the two curves $\overline{C}(n,\lambda)$ and $\overline{C}(n+1,\lambda)$ intersect at a unique point $\lambda_0$.
\vspace{-10pt}
\section{Proof of Theorem \ref{finalresultslambda}}
\label{prooftheorem3}
To show that $n(\lambda_{n_0})=n_0$, it is sufficient to show that for any $n\neq n_0$, we have that $\overline{C}(n,\lambda_{n_0})\geq \overline{C}(n_0,\lambda_{n_0})$. To prove this, the first step of our analysis consists of studying the behavior of the intersection points $\lambda(n)$ as $n$ increases. More precisely, we consider the sequence $\big(\lambda(n)\big)_{n\in\mathbb{N}}$ as the intersection point between $\overline{C}(n,\lambda)$ and $\overline{C}(n+1,\lambda)$. By using the definition in (\ref{deflambda}), we have that:
\begin{equation}
\lambda(n)=\frac{\overline{C}(n+1)-\overline{C}(n)}{A(n)-A(n+1)} \quad \forall n\in\mathbb{N}
\end{equation}
To pursue our analysis, we provide key results on the behavior of the intersection points in the following proposition.
\begin{proposition}
The sequence $\big(\lambda(n)\big)_{n\in\mathbb{N}}$ is increasing with $n$.
\label{propositionlambdaincrease}
\end{proposition}
\vspace{-10pt}
\begin{IEEEproof}
As a first step in the proof, we recall that due to the results of Lemma \ref{lemmacnincreasing} and the decreasing nature of $A(n)$, we have that $\lambda(n)\geq0\:\:\forall n\in\mathbb{N}$. As $\overline{C}(0)=\overline{C}(1)$, we can deduce that $\lambda(0)=0$ and therefore, we can restrict ourselves to study the increasing property of $\big(\lambda(n)\big)_{n\in\mathbb{N}}$ solely for the case where $n\in\mathbb{N}^*$. To that extent, as seen in Theorem \ref{theoremcostgamma}, the expression of the average cost is far from trivial. Consequently, to be able to study the variations of $\big(\lambda(n)\big)_{n\in\mathbb{N}^{*}}$, we first provide a lemma that will be useful to our analysis.
\begin{lemma}
The series $\big(\pi_0(n)\big)_{n\in\mathbb{N}^{*}}$ is decreasing with $n$.
\label{pidecreasing}
\end{lemma}
\begin{IEEEproof}
To prove this, let us consider the series $h(n)=\frac{1}{\pi_0(n+1)}-\frac{1}{\pi_0(n)} \:\:\forall n\in\mathbb{N}^{*}$. By replacing $\pi_0(n)$ and $\pi_0(n+1)$ by their respective values, we can show that:
\begin{equation}
h(n)=(N-1)p_tb^{n-1}\big(\frac{b-a}{1-a}\big).
\label{expressionhn}
\end{equation}
In other words, the series $h(n)\frac{(1-a)}{(N-1)p_t(b-a)}$ is a geometric series with a common ratio $b$. As $a<1$, we can conclude that the sign of $h(n)$ depends on the sign of $b-a$. To that extent, and by keeping in mind that $p_t<p_R$, we have that $b-a=p_R(1-p_f)-p_sp_t=p_s(p_R-p_t)>0$. Hence, we can conclude that $h(n)=\frac{\pi_0(n)-\pi_0(n+1)}{\pi_0(n)\pi_0(n+1)}\geq0 \:\:\forall n\in\mathbb{N}^{*}$. Baring in mind that $\pi_0(n)\geq0 \:\:\forall n\in\mathbb{N}^{*}$, we can assert that $\pi_0(n)\geq \pi_0(n+1) \:\:\forall n\in\mathbb{N}^{*}$ which concludes our proof.
\end{IEEEproof}
With the above lemma being laid out, we now find an explicit expression of the following difference: $\Delta\overline{C}=\overline{C}(n+1)-\overline{C}(n)$. As we have previously mentioned, the expression of the average cost is complicated, which makes treating the difference $\Delta\overline{C}$ a challenging task. To that extent, we provide in the following the $8$ terms that make up $\Delta\overline{C}$:
\begin{itemize}
\item $z_1=\frac{(N-1)p_t\big(\pi_0(n+1)-\pi_0(n)\big)}{(1-b)^2}$
\item $z_2=\frac{(N-1)p_tnb^2\big(b^n\pi_0(n+1)-b^{n-1}\pi_0(n)\big)}{(1-b)^2}$
\item $z_3=\frac{(N-1)p_tnb\big(-b^n\pi_0(n+1)+b^{n-1}\pi_0(n)\big)}{(1-b)^2}$
\item $z_4=\frac{(N-1)p_tb^{n+1}(b-1)\pi_0(n+1)}{(1-b)^2}$
\item $z_5=\frac{(N-1)p_tan\big(b^{n}\pi_0(n+1)-b^{n-1}\pi_0(n)\big)}{1-a}$
\item $z_6=\frac{(N-1)p_tab^{n}\pi_0(n+1)}{(1-a)}$
\item $z_7=\frac{(N-1)p_t\big(-b^{n+1}\pi_0(n+1)+b^{n}\pi_0(n)\big)}{(1-b)^2}$
\item $z_8=\frac{(N-1)p_ta\big(b^{n}\pi_0(n+1)-b^{n-1}\pi_0(n)\big)}{(1-a)^2}$
\end{itemize}
Next, we divide each term by the expression $A(n)-A(n+1)$ previously reported in Section V-D. By replacing the terms with their values, and after algebraic manipulations, we can verify that the terms that constitute the expression of $\lambda(n)$ are:
\begin{itemize}
\item $g_1=\frac{z_1}{A(n)-A(n+1)}=\frac{(1-a)(-b(N-1)p_t+\frac{(N-1)p_ta(1-b)}{1-a})}{(1-b)^3(1+\frac{(N-1)p_t}{1-b})}$
\item $g(n)=\frac{\sum\limits_{i=2}^{6}z_i}{A(n)-A(n+1)}=\frac{b-a}{1-b}(n-\frac{b}{\frac{\pi_0(n)}{\pi_0(n+1)}-b})$
\item $g_7=\frac{z_7}{A(n)-A(n+1)}=\frac{b(1-a)}{(1-b)^2}$
\item $g_8=\frac{z_8}{A(n)-A(n+1)}=\frac{-a}{1-a}$
\end{itemize}
We can see that $g_1$, $g_7$, and $g_8$ are only constant terms. On the other hand, the term $g(n)$ requires further investigation. To that extent, we provide the following lemma.
\begin{lemma}
The series $\big(g(n)\big)_{n\in\mathbb{N}^{*}}$ is increasing with $n$. 
\end{lemma}
\begin{IEEEproof}
First of all, let us define the ratio $r(n)$ as $\frac{\pi_0(n)}{\pi_0(n+1)}$. To study the variations of $\big(g(n)\big)_{n\in\mathbb{N}^{*}}$, we consider the difference $\Delta g(n)=g(n+1)-g(n)$. By using the expression of $g(n)$, we have that:
\begin{equation}
\Delta g(n)=\frac{r(n)\big(r(n+1)-2b\big)+b^2}{\big(r(n+1)-b\big)\big(r(n)-b\big)}.
\label{deltagn}
\end{equation}
As $r(n)\geq1\geq b \:\:\forall n\in\mathbb{N}^{*}$ (we recall the results of Lemma \ref{pidecreasing}), we can conclude that it is enough to study the sign of the numerator in (\ref{deltagn}). By replacing $r(n)$ with its expression, we can see that to prove $\Delta g(n)\geq0$, it is sufficient to have:
\begin{equation}
\frac{2b}{\pi_0(n+1)}-\frac{1}{\pi_0(n+2)}-\frac{b^2}{\pi_0(n)}\leq0.
\label{pinnplus1}
\end{equation}
By replacing $\pi_0(n),\pi_0(n+1)$ and $\pi_0(n+2)$ with their expressions using (\ref{pi0}), we can show that the LHS of (\ref{pinnplus1}) becomes $-(b-1)^2(1+\frac{(N-1)p_t}{1-b})$ which is always negative since $b\leq1$. Therefore, we have that $\big(g(n)\big)_{n\in\mathbb{N}^{*}}$ is an increasing sequence with $n$.
\end{IEEEproof}
From the above lemma, we can conclude that $\lambda(n)$ is the sum of two terms: a constant and an increasing function with $n$. Therefore, the sequence $\big(\lambda(n)\big)_{n\in\mathbb{N}^{*}}$ is increasing with $n$ which concludes our proof.
\end{IEEEproof}
Our subsequent analysis will be divided into two sections where we study the thresholds $n$ that are larger than $n_0$ and prove that they lead to a cost $\overline{C}(n,\lambda_{n_0})$ that is higher than $\overline{C}(n_0,\lambda_{n_0})$. The case where $n<n_0$ is then tackled in the section after it.
\subsubsection{$n>n_0$}
\begin{figure}[!ht]
\centering
\includegraphics[width=.75\linewidth]{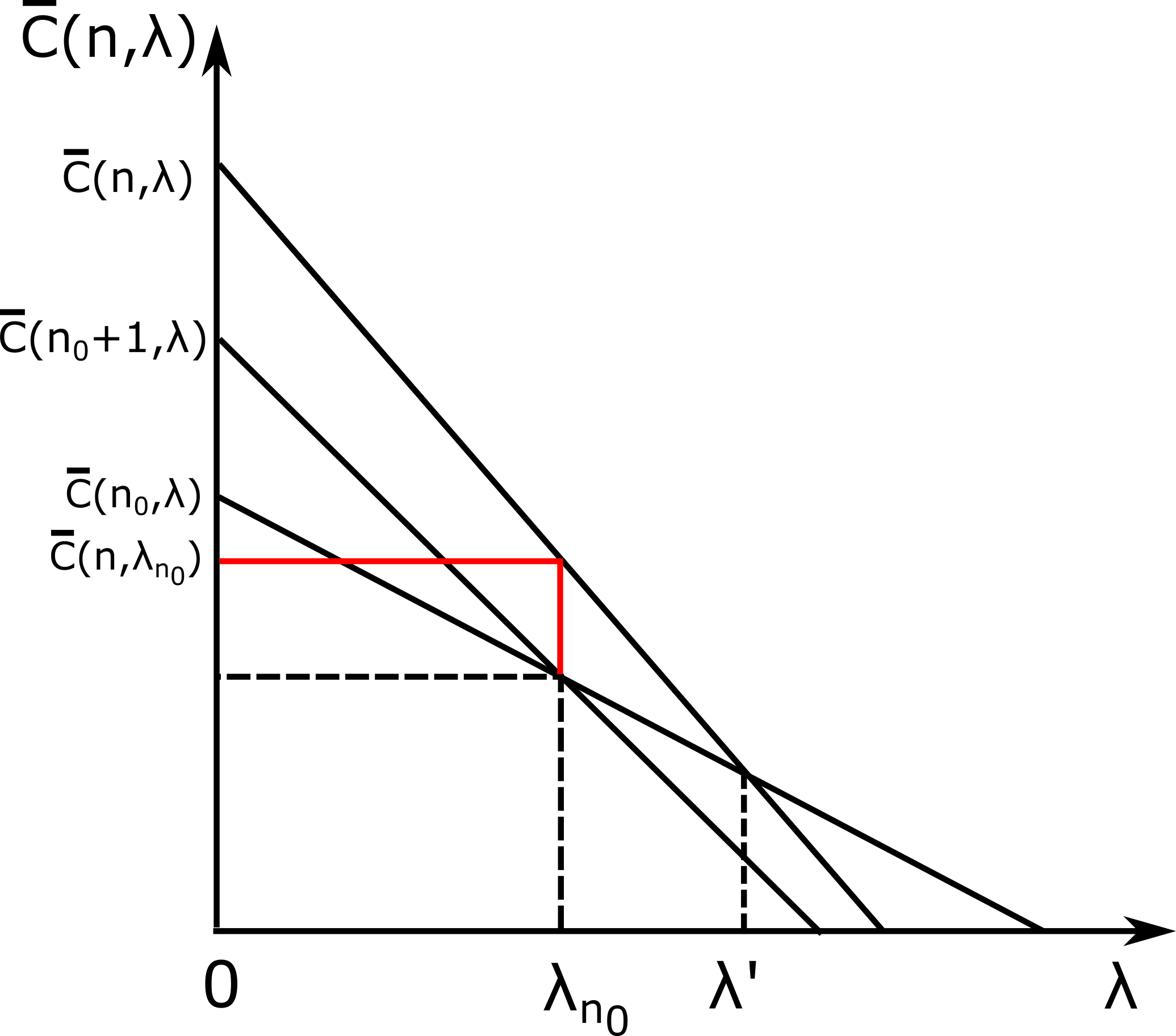}
\setlength{\belowcaptionskip}{-5pt}
\caption{Illustration of the proof: $n>n_0$.}
\label{prooflambdaplusgrand}
\end{figure}

To analyze this case, we first provide the following lemma.
\begin{lemma}
$\forall k_2>k_1$, we consider two sequences $\big(U_1(n)\big)_{n\in\mathbb{N}}$ and $\big(U_2(n)\big)_{n\in\mathbb{N}}$ such that $\big(U_2(n)\big)_{n\in\mathbb{N}^{*}}$ is an increasing sequence. If $\frac{U_1(n+1)-U_1(n)}{U_2(n+1)-U_2(n)}$ increases with $n$, then the following holds:
\begin{equation}
\frac{U_1(k_2)-U_1(k_1)}{U_2(k_2)-U_2(k_1)}\geq \frac{U_1(k_1+1)-U_1(k_1)}{U_2(k_1+1)-U_2(k_1)}.
\end{equation}
\label{lemmasaad1}
\end{lemma}
\vspace{-15pt}
\begin{IEEEproof}
The proof is based on mathematical induction. More precisely, we know that the above lemma is true for $k_2=k_1+1$. We suppose that it is true for any $k_2>k_1+1$ and investigate the property for $k_2+1$. To that extent, we have that $\frac{U_1(k_2+1)-U_1(k_1)}{U_2(k_2+1)-U_2(k_1)}$ can be rewritten as:
\begin{equation}
\frac{U_1(k_2+1)-U_1(k_2)}{U_2(k_2+1)-U_2(k_1)}+\frac{U_1(k_2)-U_1(k_1)}{U_2(k_2+1)-U_2(k_1)}.
\end{equation}
By multiplying the first and second term by $\frac{U_2(k_2+1)-U_2(k_2)}{U_2(k_2+1)-U_2(k_2)}$ and $\frac{U_2(k_2)-U_2(k_1)}{U_2(k_2)-U_2(k_1)}$ respectively, and by taking into account the increasing property of the ratio $\frac{U_1(n+1)-U_1(n)}{U_2(n+1)-U_2(n)}$ along with the induction assumption, the results can be found to be true for $k_2+1$ which concludes our proof.
\end{IEEEproof}
We can apply the above lemma by taking $U_1(n)=\overline{C}(n)$,  $U_2(n)=-A(n)$ and noting the results of Proposition \ref{propositionlambdaincrease} on $\big(\lambda(n)\big)_{n\in\mathbb{N}}$. Consequently, Lemma \ref{lemmasaad1} tell us that the intersection between $\overline{C}(n,\lambda)$ and $\overline{C}(n_0,\lambda)$ for any $n>n_0+1$ occur after $\lambda_{n_0}$. By observing Fig. \ref{prooflambdaplusgrand}, we can see that this leads to $\overline{C}(n,\lambda_{n_0})$ being larger than $\overline{C}(n_0,\lambda_{n_0})$ due to the properties of the curve $\overline{C}(n,\lambda)$ previously reported in Lemma \ref{lemmacnincreasing}.
\subsubsection{$n<n_0$} 
\begin{figure}[!ht]
\centering
\includegraphics[width=.74\linewidth]{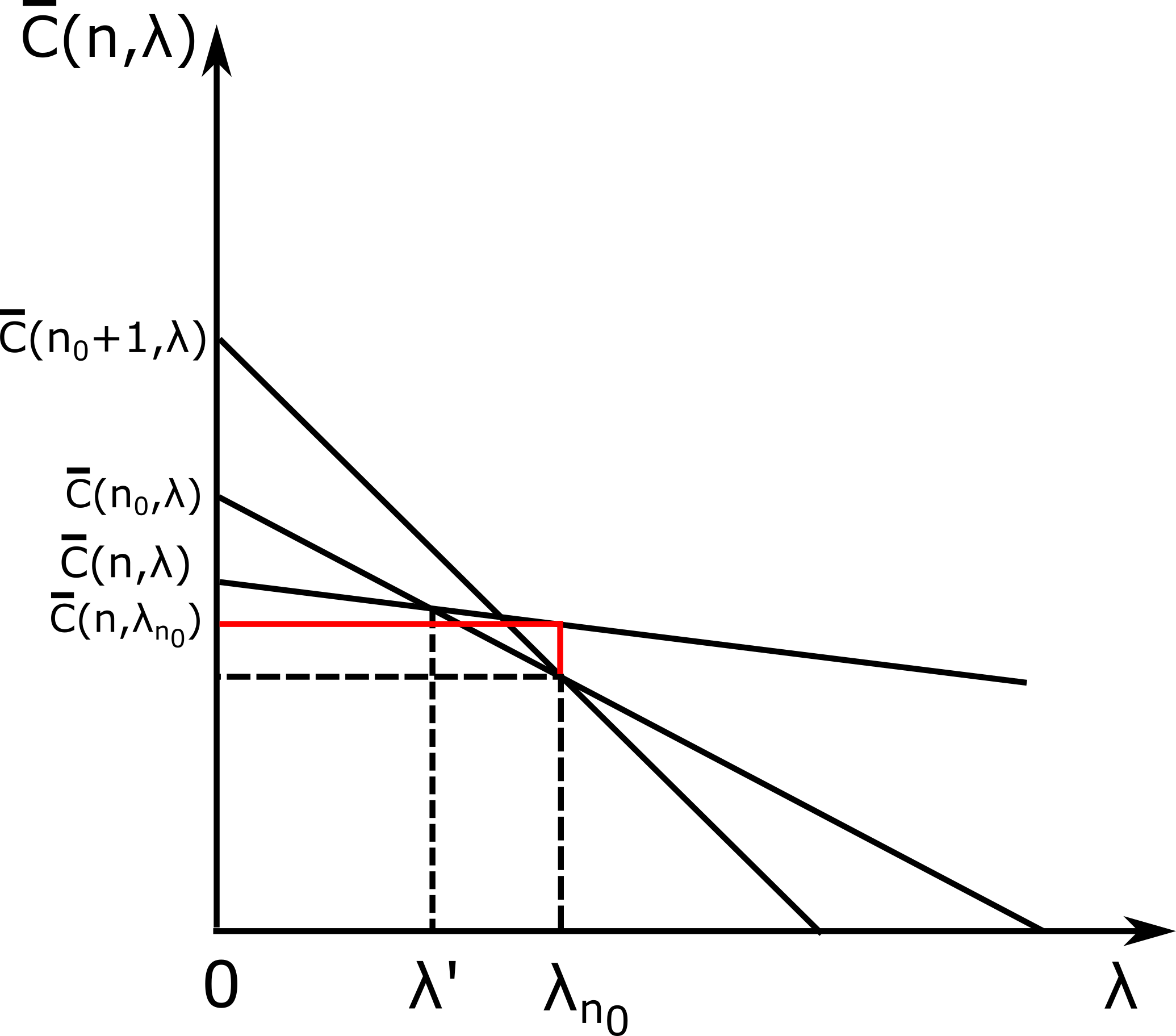}
\setlength{\belowcaptionskip}{-15pt}
\caption{Illustration of the proof: $n<n_0$.}
\label{prooflambdapluspetit}
\end{figure}

Similarly to the previous subsection, we provide two vital lemmas to our analysis.
\begin{lemma}
$\forall n\geq 1$, if the conditions of Lemma \ref{lemmasaad1} are satisfied, we have that:
\begin{equation}
\frac{U_1(n)-U_1(n-1)}{U_2(n)-U_2(n-1)}\leq \frac{U_1(n+1)-U_1(n-1)}{U_2(n+1)-U_2(n-1)},
\end{equation}
\begin{equation}
\frac{U_1(n+1)-U_1(n-1)}{U_2(n+1)-U_2(n-1)} \leq \frac{U_1(n+1)-U_1(n)}{U_2(n+1)-U_2(n)}.
\end{equation}
\label{lemmasaad2}
\end{lemma}
\vspace{-20pt}
\begin{IEEEproof}
We first start by rewriting $\frac{U_1(n+1)-U_1(n-1)}{U_2(n+1)-U_2(n-1)}$ as $\frac{U_1(n+1)-U_1(n)}{U_2(n+1)-U_2(n-1)}+\frac{U_1(n)-U_1(n-1)}{U_2(n+1)-U_2(n-1)}$. Afterward, the proof is based on multiplying the above expression by $\frac{U_2(n+1)-U_2(n)}{U_2(n+1)-U_2(n)}$ and $\frac{U_2(n)-U_2(n-1)}{U_2(n)-U_2(n-1)}$ and using the conditions of the lemma to prove the LHS and RHS inequalities, respectively. The details are omitted for the sake of space.
\end{IEEEproof}
\begin{lemma}
$\forall n\leq n_0-1$, we always have that:
\begin{equation}
\frac{\overline{C}(n_0)-\overline{C}(n)}{A(n)-A(n_0)}\geq \frac{\overline{C}(n_0)-\overline{C}(n-1)}{A(n-1)-A(n_0)}\geq \frac{\overline{C}(n)-\overline{C}(n-1)}{A(n-1)-A(n)}.
\end{equation}
\label{finallemmasaad2}
\end{lemma}
\vspace{-15pt}
\begin{IEEEproof}
The proof is based on a mathematical \emph{backward} induction. As a first step, we tackle the case for $n=n_0-1$. As $\lambda(n)$ is increasing with $n$, we have that $\frac{\overline{C}(n_0)-\overline{C}(n_0-1)}{A(n_0-1)-A(n_0)}\geq\frac{\overline{C}(n_0-1)-\overline{C}(n_0-2)}{A(n_0-2)-A(n_0-1)}$. By applying Lemma \ref{lemmasaad2} for $n=n_0-1$, we can conclude that the above property is true for $n=n_0-1$. We now suppose that this property holds for any $n<n_0-1$ and aim to prove it to be true for $n-1$. By using our supposition, along with the increasing property of $\lambda(n)$ and the results of Lemma \ref{lemmasaad2}, the property can be verified to be true for $n-1$ which concludes our proof. 
\end{IEEEproof}
Equipped with the above two lemmas, we will be able to show that for any $n<n_0$, we have that $\overline{C}(n_0,\lambda_{n_0})\leq \overline{C}(n,\lambda_{n_0})$. To do so, we aim to show that the intersection between the curves $\overline{C}(n,\lambda)$ and $\overline{C}(n_0,\lambda)$ for any $n<n_0$ occur before $\lambda_{n_0}$. Combined with the properties of the curve $\overline{C}(n,\lambda)$ previously reported in Lemma \ref{lemmacnincreasing}, we can see in Fig. \ref{prooflambdapluspetit} that this is equivalent to what we are aiming to prove. Our goal is, therefore, summarized in proving that: $\frac{\overline{C}(n_0+1)-\overline{C}(n_0)}{A(n_0)-A(n_0+1)}\geq \frac{\overline{C}(n_0)-\overline{C}(n)}{A(n)-A(n_0)}$ for any $n<n_0$. From the first inequality of the results of Lemma \ref{finallemmasaad2}, we can conclude that the series $\frac{\overline{C}(n_0)-\overline{C}(n)}{A(n)-A(n_0)}$ is increasing with $n$ for all $n\leq n_0-1$. Therefore, we have that for all $n<n_0$:
\begin{equation}
\frac{\overline{C}(n_0)-\overline{C}(n_0-1)}{A(n_0-1)-A(n_0)}\geq\frac{\overline{C}(n_0)-\overline{C}(n)}{A(n)-A(n_0)}.
\label{ekherequation}
\end{equation}
Lastly, by using the fact that $\lambda(n)$ is increasing with $n$, we can conclude that: $\frac{\overline{C}(n_0+1)-\overline{C}(n_0)}{A(n_0)-A(n_0+1)}\geq\frac{\overline{C}(n_0)-\overline{C}(n_0-1)}{A(n_0-1)-A(n_0)}$. Combining this with the results of eq. (\ref{ekherequation}), we can conclude our proof.

\newpage
\clearpage
\setcounter{page}{1}
\begin{strip}
\vspace{-30pt}
   \begin{center}
      \Huge Supplementary Material for the paper ``The Age of Incorrect Information: A New Performance Metric for Status Updates" \\
   \end{center}

\end{strip}
\appendices
\setcounter{section}{2}
\section{Proof of Lemma \ref{increasingvalue}}
\label{prooflemma1}
Our proof is based on the well-known value iteration algorithm (\textbf{VIA}) \cite{Bertsekas:2000:DPO:517430}. By letting $V_t(.)$  be the value function at iteration $t$, the VIA consists of updating the value function as follows:
\begin{equation}
V_{t+1}(S)=\min_{\psi\in\{0,1\}}\big\{S+\sum_{S'\in\mathbb{N} }\Pr(S\rightarrow S'|\psi)V_{t}(S')\big\} \quad \forall S\in\mathbb{N}.
\label{VIA}
\end{equation}
Regardless of the initial value $V_0(S)$, it is well-known that the algorithm converges to the value function of the Bellman equation (\ref{bellman}) \cite{Bertsekas:2000:DPO:517430} (i.e., $\lim_{t\to+\infty} V_{t}(S)=V(S) \:\:\forall S\in\mathbb{N}$). Consequently, to infer on the monotonicity of $V(S)$, it is sufficient to prove that $\forall S_2\geq S_1$:
\begin{equation}
V_t(S_2)\geq V_t(S_1) \quad t=0,1,\ldots
\label{conditionmonotone}
\end{equation}
To proceed in that direction, and without loss of generality, we suppose that $V_0(S)=0\:\:\forall S\in\mathbb{N}$. Therefore, (\ref{conditionmonotone}) holds for $t=0$. Next, we suppose that  the condition in (\ref{conditionmonotone}) is true up till $t>0$ and we examine if it holds for $t+1$. To do so, we examine the Right Hand Side (RHS) of (\ref{VIA}) for both states $S_2$ and $S_1$. To that extent, we first take the case where $S_1\neq0$ and we distinguish between the two possible transmission decisions $\psi$:
\begin{itemize}
\item $\psi=0$: In this case, the RHS is equal to $x=S_1+(p_R+(N-2)p_t)V_t(S_1+1)+p_tV_t(0)$ and $y=S_2+(p_R+(N-2)p_t)V_t(S_2+1)+p_tV_t(0)$ for $S_1$ and $S_2$ respectively. Baring in mind that $V_t(S_2)\geq V_t(S_1)$, we can easily see that $x\leq y$.
\item $\psi=1$: In this case, the RHS is equal to $z=S_1+(p_Rp_f+(N-2)p_t+p_sp_t)V(S_1+1)+(p_Rp_s+p_fp_t)V(0)$ and $w=S_2+(p_Rp_f+(N-2)p_t+p_sp_t)V(S_2+1)+(p_Rp_s+p_fp_t)V(0)$ for $S_1$ and $S_2$ respectively. Taking into account that $V_t(S_2)\geq V_t(S_1)$, we can also verify that $z\leq w$. 
\end{itemize}
Lastly, we know that if $x\leq y$ and $z\leq w$ then $\min(x,z)\leq\min(y,w)$. For the case where $S_1=0$, we can show that $x=z=p_RV_t(0)+(N-1)p_tV_t(1)$. After some algebraic manipulations, we can easily verify that the same above inequalities still holds. Consequently, we can assert that $V_{t+1}(S_2)\geq V_{t+1}(S_1) \:\:\forall t,S_1,S_2\in\mathbb{N}$. This concludes our inductive proof that shows that the value function $V(S)$ is increasing in $S \:\:\forall S\in\mathbb{N}$.
\section{Proof of Theorem \ref{optimalpolicyunconstrained}}
\label{prooftheorem1}
As we have previously stated, it is well-known that the optimal transmission policy can be obtained by solving the Bellman equation in (\ref{bellman}). On top of that, we recall that the VIA, previously reported in the proof of Lemma \ref{increasingvalue}, converges to the value function of the Bellman equation in (\ref{bellman}). Consequently, we can deduce the optimal sequence of actions based on the value function at each time instant $t$ by reconsidering the VIA:
\begin{equation}
V_{t+1}(S)=\min_{\psi\in\{0,1\}}\big\{S+\sum_{S'\in\mathbb{N} }\Pr(S\rightarrow S'|\psi)V_{t}(S')\big\} \quad \forall S\in\mathbb{N}
\end{equation}
To that extent, let us define $\Delta V_{t+1}(S)$ as the difference between the value functions if the transmitter sends a packet or remains idle for any state $S$. More specifically, we have that $\Delta V_{t+1}(S)=V^{1}_{t+1}(S)-V^{0}_{t+1}(S)$ where $V^{1}_{t+1}(S)$ and $V^{0}_{t+1}(S)$ are the value functions at time $t+1$ if $\psi=1$ and $\psi=0$ respectively. By obeying to the dynamics reported in Section III-B, we have:
\begin{equation}
\Delta V_{t+1}(0)=0,
\end{equation}
\begin{equation}
\Delta V_{t+1}(S)=p_s(p_t-p_R)(V_t(S+1)-V_t(0)) \quad \forall S\in\mathbb{N}^{*}.
\end{equation}
The first thing we see is that when the state of the system is $S(t)=0$, both actions of remaining idle or transmitting leads to the same value function at time $t+1$. We can now tackle the case where $S(t)\neq0$. To that extent, and as $V_t(S)$ is always increasing with $S$ (Lemma \ref{increasingvalue}), we can assert that $(V_t(S+1)-V_t(0))\geq0$. Based on this, we distinguish between the following cases:
\subsubsection{$p_t<p_R$} 
\begin{figure}[!ht]
\centering
\includegraphics[width=.9\linewidth]{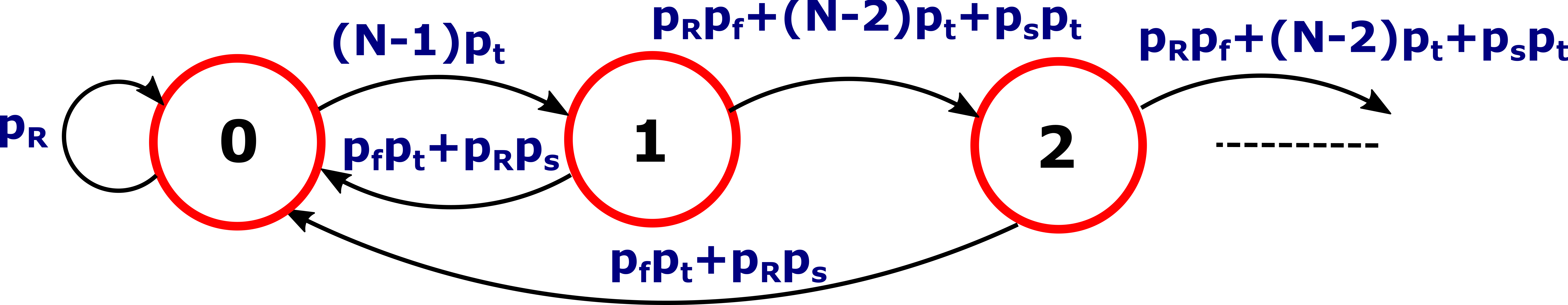}
\setlength{\belowcaptionskip}{-5pt}
\caption{The states transitions under the ``always update" policy.}
\label{alwaystransmitdtmc}
\end{figure}

In this scenario, we can see that $\Delta V_{t+1}(S)$ is always negative for any $S\neq0$. Consequently, it is always optimal to transmit a packet when $S(t)\neq0$. Combined with the fact that a transmission or remaining idle leads to the same value function when $S(t)=0$, we can conclude that the optimal policy is to either send updates at each time slot or send updates when the receiver is in an erroneous state (i.e., when $S(t)\neq0$). To calculate the average cost in this case, we can see that in the case of an ``always update" policy, the MDP can be modeled through a Discrete Time Markov Chain (\textbf{DTMC}) where:
\begin{itemize}
\item The states refer to the values of the penalty function $S(t)$.
\item The dynamics of $S(t) \:\:\forall (S,t)$ coincide with those of $\psi(t)=1$ of Section III-B. 
\end{itemize}
The DTMC mentioned above is reported in Fig. \ref{alwaystransmitdtmc}.
\\
To find the average cost in this case, we first provide the following lemma.
\begin{lemma}
The DTMC of the ``always update" policy is irreducible and admits $\pi_k\:\: \forall k\in\mathbb{N}$ as its stationary distribution where:
\begin{equation}
\pi_0=\frac{1}{1+\frac{(N-1)p_t}{1-a}},
\label{alwaysupdatepi0}
\end{equation}
\begin{equation}
\pi_k=(N-1)p_ta^{k-1}\pi_0 \quad k\geq1,
\label{alwaysupdatepik}
\end{equation}
with the constant $a$ being equal to $p_Rp_f+(N-2)p_t+p_sp_t$.
\label{alwaysupdatestationary}
\end{lemma}
\begin{IEEEproof}
It is sufficient to formulate the general balance equations at any state $k\geq2$, which leads to $\pi_k=a\pi_{k-1}$. By proceeding with a forward induction, and knowing that $\pi_1=(N-1)p_t\pi_0$, the results of (\ref{alwaysupdatepik}) can be found. Next, by taking into account the fundamental equality $\sum\limits_{k=0}^{+\infty} \pi_k=1$, we can find $\pi_0$ which concludes our proof.
\end{IEEEproof}
To find the average cost of the above DTMC, we first note that the cost incurred by being at state $S=k$ is nothing but the value $k$ of the state itself. Consequently, we have that $\overline{C}_{AU}=\sum\limits_{k=1}^{+\infty}k\pi_k$. By taking into account the above stationary distribution, and the following series equalities, the expression in (\ref{alwaysupdate}) can be found.
\begin{equation}
\sum\limits_{k=1}^{+\infty}a^{k-1}=\frac{1}{1-a}, \quad\quad \sum\limits_{k=1}^{+\infty}ka^{k-1}=\frac{1}{(1-a)^2}.
\end{equation}
\subsubsection{$p_t\geq p_R$} 
\begin{figure}[!ht]
\centering
\includegraphics[width=.9\linewidth]{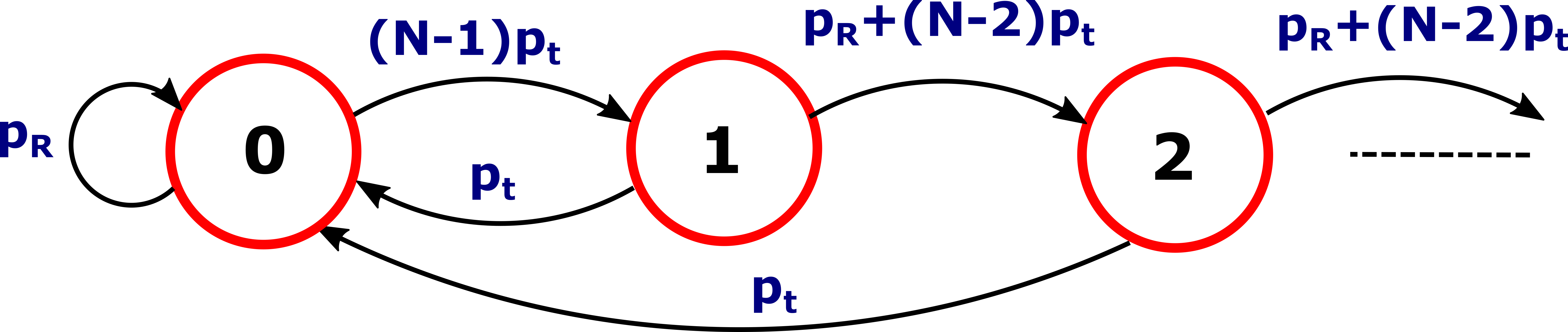}
\setlength{\belowcaptionskip}{-5pt}
\caption{The states transitions under the ``never transmit" policy.}
\label{nevertransmitdtmc}
\end{figure}

In this case, we can see that $\Delta V_{t+1}(S)$ is always positive. Combined with the fact that a transmission or remaining idle leads to the same value function when $S(t)=0$, we can conclude that the optimal policy is always to remain idle. The intuition behind this is that when $p_t\geq p_R$, any packet being transmitted about the information source has a high chance of becoming obsolete by the time it reaches the monitor. To calculate the average cost in the case where the transmitter is always idle, the MDP can be modeled through the DTMC reported in Fig. \ref{nevertransmitdtmc}.
\\
The analysis of the above DTMC is the same as the one of the previous case ($p_t<p_R$). More specifically, it is sufficient to substitute $a$ by $b$ where $b=p_R+(N-2)p_t$ in (\ref{alwaysupdate}) to obtain the expression in (\ref{noupdatescost}).
\section{Proof of Proposition \ref{propositionthreshold}}
\label{proofproposition1}
The proof follows the same direction as that of Theorem \ref{optimalpolicyunconstrained}. More precisely, the optimal transmission policy can be obtained by solving the Bellman equation formulated in (\ref{bellmannew}). To that extent, we leverage the VIA to find the optimal stransmission sequence. In other words, and as it has been done before, we investigate $\Delta V_{t+1}(S)=V^{1}_{t+1}(S)-V^{0}_{t+1}(S)$ where $V^{1}_{t+1}(S)$ and $V^{0}_{t+1}(S)$ are the value functions at time $t+1$ if $\psi=1$ and $\psi=0$ respectively. By obeying to the dynamics reported in Section III-B, we have:
\begin{equation}
\Delta V_{t+1}(0)=\lambda,
\end{equation}
\begin{equation}
\Delta V_{t+1}(S)=\lambda+p_s(p_t-p_R)(V_t(S+1)-V_t(0)) \quad \forall S\in\mathbb{N}^{*}.
\end{equation}
As $\lambda\geq0$, we can conclude that the action of remaining idle is always optimal when $S=0$. As for the case where $S\neq0$, we can see that $\Delta V_{t+1}(S)$ is the sum of a positive constant and a decreasing non-positive function. Consequently, we have that the optimal action is increasing with $S$ from $\psi^*=0$ to $\psi^*=1$. In other words, the difference $\Delta V_{t+1}(S)$ decreases with $S$, and at a certain point, the action of transmitting becomes more beneficial than remaining idle. Therefore, we can conclude that the optimal policy of the problem is of a threshold nature.
\section{Proof of Proposition \ref{stationarydistribution}}
\label{proofproposition2}
To proceed with the proof, we first formulate the general balance equation at state $1$ which leads to $\pi_1(n)=(N-1)p_t\pi_0(n)$. Afterward, we provide the general balance equations at states $k$, with $2\leq k\leq n$:
\begin{equation}
\pi_k(n)=\big(p_R+(N-2)p_t\big)\pi_{k-1}(n) \quad \quad 2\leq k\leq n
\end{equation}
By noting the results above, along with those on $\pi_1(n)$, and by carrying on with a forward induction, the results of (\ref{pik}) can be found. Next, we formulate the balance equations at states $k$, with $k\geq n+1$:
\begin{equation}
\pi_k(n)=\big(p_Rp_f+(N-2)p_t+p_sp_t\big)\pi_{k-1}(n) \quad   k\geq n+1
\end{equation}
By using the above results, and those of (\ref{pik}), and by proceeding with a forward induction, we can find the equations in (\ref{pin}). Lastly, we make use of the following fundamental equality:
\begin{equation}
\sum\limits_{k=0}^{+\infty} \pi_k(n)=1.
\label{sumequal1}
\end{equation}
By replacing $\pi_k(n)$ with their values in (\ref{sumequal1}) and by noting the following series results:
\begin{equation}
\sum\limits_{k=1}^{n}b^{k-1}=\frac{1-b^n}{1-b},
\end{equation}
\begin{equation}
\sum\limits_{k=n+1}^{+\infty}a^{k-n}=\frac{a}{1-a},
\end{equation}
we can find $\pi_0(n)$ which concludes our proof.
\section{Proof of Proposition \ref{theoremcostgamma}}
\label{prooftheorem2}
To calculate the average cost of the threshold policy, we first note that the cost incurred by being at state $S=k$ is nothing but the value $k$ of the state itself. Moreover, the transmitter attempts to send a packet solely when $S\geq n$. Consequently, we have that $\overline{C}(n,\lambda)=\overline{C}(n)+\overline{C}_1(n,\lambda)$ where: 
\begin{equation}
\overline{C}(n)=\sum\limits_{k=1}^{+\infty}k\pi_k(n),
\end{equation}
\begin{equation}
\overline{C}_1(n,\lambda)=\lambda\sum\limits_{k=n}^{+\infty}\pi_k-\lambda\alpha.
\end{equation}
By replacing $\pi_k(n)$ with its value from Proposition \ref{stationarydistribution}, we have that:
\begin{equation}
\overline{C}(n)=(N-1)p_t\pi_0\big(\sum\limits_{k=1}^{n}kb^{k-1}+\sum\limits_{k=n+1}^{+\infty}b^{n-1}ka^{k-n}\big).
\label{averagecostexpression}
\end{equation}
To further simplify the above expression, we first note that the series $\sum\limits_{k=1}^{n}kb^{k-1}$ is nothing but the derivative with respect to $b$ of the series $\sum\limits_{k=0}^{n}b^{k}=\frac{1-b^{n+1}}{1-b}$. Consequently, by deriving the expression in the right hand side, we have that:
\begin{equation}
\sum\limits_{k=1}^{n}kb^{k-1}=\frac{1+b^n(nb-n-1)}{(1-b)^2}.
\end{equation}
Next, we can address the second term of the expression in (\ref{averagecostexpression}). To that extent, we proceed with a change of variables $k'=k-n$. With that being done, and by noting the fact that $\sum\limits_{k'=1}^{+\infty}k'a^{k'}=\frac{a}{(1-a)^2}$, the expression in (\ref{costgamma}) can be found. By pursuing the same series analysis, we can deduce the expression in (\ref{costlambda}) which concludes our proof.
\end{document}